\documentclass[final,3p,times]{elsyearticle}

\usepackage{graphicx}
\usepackage{sidecap}

\usepackage{amssymb}

\usepackage[english,francais]{babel}
\usepackage[utf8]{inputenc}
\usepackage[T1]{fontenc}

\def\og{\leavevmode\raise.3ex\hbox{$\scriptscriptstyle\langle\!\langle$~}}
\def\fg{\leavevmode\raise.3ex\hbox{~$\!\scriptscriptstyle\,\rangle\!\rangle$}}
\def\og{\lq\lq}
\def\fg{\rq\rq}

\usepackage[colorinlistoftodos]{todonotes}
\usepackage[colorlinks=true, allcolors=blue]{hyperref}
\usepackage{braket}
\usepackage{subfig}
\usepackage{bbold}
\usepackage{amsmath}
\usepackage{graphicx}
\usepackage{natbib}
\usepackage{wrapfig}
\usepackage{comment}

\newcommand{\bea}{\begin{eqnarray}}
\newcommand{\eea}{\end{eqnarray}}
\newcommand{\be}{\begin{equation}}
\newcommand{\ee}{\end{equation}}

\begin{document}

\selectlanguage{english}
\begin{frontmatter}



\title{Squeezing of nonlinear spin observables by one axis twisting in the presence of decoherence: An analytical study}


\author[ad1]{Youcef Baamara}
\author[ad1]{Alice Sinatra}
\author[ad1,ad2]{Manuel Gessner}

\address[ad1]{Laboratoire Kastler Brossel, ENS-Universit\'e PSL, CNRS, Universit\'e de la Sorbonne and Coll\`ege de France, 24 rue Lhomond, 75231 Paris, France}
\address[ad2]{ICFO-Institut de Ci\`{e}ncies Fot\`{o}niques, The Barcelona Institute of Science and Technology, Av. Carl Friedrich Gauss 3, 08860, Castelldefels (Barcelona), Spain}

\begin{abstract}
\noindent
{ In an ensemble of two-level atoms that can be described in terms of a collective spin, entangled states can be used to enhance the sensitivity of interferometric precision measurements. While non-Gaussian spin states can produce larger quantum enhancements than spin-squeezed Gaussian states, their use requires the measurement of observables that are nonlinear functions of the three components of the collective spin. In this paper we develop strategies that achieve the optimal quantum enhancements using non-Gaussian states produced by a nonlinear one-axis-twisting Hamiltonian, and show that measurement-after-interaction techniques, known to amplify the output signals in quantum parameter estimation protocols, are effective in measuring nonlinear spin observables. Including the presence of the relevant decoherence processes from atomic experiments, we determine analytically the quantum enhancement of non-Gaussian over-squeezed states as a function of the noise parameters for arbitrary atom numbers.}

\noindent 
\vskip 0.5\baselineskip
\selectlanguage{english}
\end{abstract} 
\end{frontmatter}


\section{Introduction}
The classical precision limit of interfermetric measurements is determined by quantum projection noise. Entangled many-body spin states with { correlated quantum fluctuations} can overcome this limit and may offer significant precision enhancements~\cite{CavesPRD1981,WinelandPRA1992,BollingerPRA1996,GiovannettiNATPHOT2011,PezzeRMP2018}. { A widely known strategy offering quantum-enhanced precision} in atomic Ramsey spectroscopy measurements is  {  spin squeezing \cite{KitagawaPRA1993,WinelandPRA1992}}: By { redistributing} the quantum noise into unmeasured observables, the variance of the spin { component} that contains the information about the phase parameter $\phi$ of interest can be reduced below the { standard quantum limit (SQL)~: $(\Delta\phi)_{\mathrm{SQL}}^2=1/N$ that is the minimum uncertainty for $N$ non-entangled atoms}. 

To generate the required quantum entanglement, well controlled interactions are used. In Bose-Einstein condensates, atomic collisions naturally generate entanglement~\cite{SorensenNATURE2001,EPJD,GrossNATURE2010,Treutlein2010}. Alternatively, { effective interactions mediated by an electromagnetic field can be implemented} in optical cavities~\cite{LerouxPRL2010}. In both cases, the one-axis-twisting (OAT) Hamiltonian $\hat{H}=\hbar\chi \hat{S}_z^2$, nonlinear in the spin component $\hat{S}_z$ where $\chi$ is determined by the interaction strength, allows for a unifying description of these interactions. 
Starting from a coherent spin state, an eigenstate of $\hat{S}_x$, the one-axis-twisting { evolution allows for} the generation of states where a linear (L) spin component, a combination of $\hat{S}_y$ and $\hat{S}_z$, is squeezed, i.e., its uncertainty is decreased below { the standard quantum limit}. Although it was shown to be an experimentally { robust} to improve measurement precision in atomic interferometers~\cite{GrossNATURE2010,Treutlein2010,LerouxPRL2010,MitchellPRL2012,HostenNATURE2016,CoxPRL2016}, this approach offers a { quantum gain that is limited to} $(\Delta\phi)_{\mathrm{SQL}}^2/(\Delta\phi)^2_{\rm L}\propto N^{2/3}$~\cite{KitagawaPRA1993}. 

One-axis-twisting generates states that are more sensitive than spin-squeezed states when the evolution is continued beyond the best linear squeezing time,  { eventually reaching the Heisenberg limit 
$(\Delta\phi)_{\mathrm{SQL}}^2/(\Delta\phi)^2_{\rm HL}= N$ that is the maximum gain allowed by quantum mechanics. 
Recently, an experiment reaching the Heisenberg limit was realized using the spin of a highly magnetic atom $2J=16$~\cite{ChalopinNCom2018}, and an experimental demonstration of a quantum gain reaching Heisenberg scaling, i.e. $(\Delta\phi)_{\mathrm{SQL}}^2/(\Delta\phi)^2_{\rm HS}= aN$ with $a<1$, was realized 
 with up to $N=350$ Ytterbium atoms~\cite{VuleticArxiv2021}.
To exploit the sensitive features of these highly entangled states, measurement-after-interaction (MAI) 
strategies such as squeezing echos have been developed \cite{YurkePRA1986,SchleierSmithPRL2016,FrowisPRL2016,MacriPRA2016,HostenSCIENCE2016,NolanPRL2017,HainePRA2018,HammererQuantum2020} that reduce the sensitivity to imperfections and detection noise.
However, their { fragility towards} decoherence~\cite{HuelgaPRL1997,MonzPRL2011,DemkowiczNATCOMMUN2012,Kittens}, and the need for stable and coherent interactions on sufficiently long time scales renders the reach of Heisenberg scaling in systems with large atom number extremely challenging.} 

A promising alternative is provided by over-squeezed spin states~\cite{StrobelSCIENCE2014,BohnetSCIENCE2016,EvrardPRL2019,XuArXiv2021} that are generated by OAT after the linear squeezing time but on time scales that are shorter than those needed to reach Heisenberg scaling. The sensitivity of these states cannot be captured in terms of the squeezing of linear spin observables, but instead requires the measurement of nonlinear spin observables~\cite{GessnerPRL2019} whose squeezing can lead to significant { quantum enhancements} beyond the reach of linear spin squeezing. { Theoretically, the metrological potential of this relevant class of states in the limit of large $N$} is only accessible by analytical approaches since numerical simulations are limited to moderate particle numbers that are too low to extrapolate the scaling behavior.

In this paper, after recalling the most important results of the squeezing of a linear (L) spin observable, we focus on the squeezing of nonlinear spin observables generated by the OAT evolution, its sensitivity enhancement beyond the linear spin squeezing and its scaling with the atom number for $N\gg 1$. First, we show that  { when a single nonlinear spin observable (NL) of the form $\{S_x,S_z\}$} is added to the linear components { in the ensemble of accessible observables, the best quantum gain} scales as $(\Delta\phi)_{\mathrm{SQL}}^2/(\Delta\phi)^2_{\rm NL}\propto N^{4/5}$ and it is reached on the time scale $\chi t\propto N^{-3/5}$; { while for an optimal linear combination of arbitrary linear and quadratic (Q) spin observables, the best quantum gain scales as $(\Delta\phi)_{\mathrm{SQL}}^2/(\Delta\phi)^2_{\rm Q}\propto N^{6/7}$ and is reached on a time scale $\chi t\propto N^{-4/7}$}. Second, we show that the { measurement-after-interaction} technique gives access to a continuous family of nonlinear spin observables { that} reproduce all the scaling laws mentioned above. More generally, we show that on time scales $\chi t\propto N^{-\alpha}$ of the { one-axis twisting} evolution with $1\geq \alpha\geq 1/2$, the MAI technique allows { one} to achieve a maximal quantum gain that scales as $(\Delta\phi)_{\mathrm{SQL}}^2/(\Delta\phi)^2_{\rm MAI}\propto N^{2-2\alpha}$. By comparing to the quantum Fisher information, which quantifies the maximal sensitivity enhancement over all possible measurements, we demonstrate that the scaling law of the MAI technique is optimal at any time in the considered time window $1 \geq \alpha \geq 1/2$. In order to study the effect of decoherence on this scaling law, we include two collective dephasing { processes} corresponding to realistic noise in atomic experiments into our analytical study: { For} a ballistic dephasing processes, described by fluctuating energy levels in the Hamiltonian, we predict a critical value of the preparation time at which we observe a discontinuous change in the scaling law { of the quantum gain}. { For} a dephasing of diffusive nature, described by a Lindblad master equation, we find that the scaling exponent is reduced by a factor of 2 independently of the dephasing strength. { In addition to the scaling laws in the large-$N$ limit, first reported in Ref.~\cite{BaamaraPRL2021}, we present general expressions of the quantum gain for arbitrary atom numbers and  identify finite-size corrections}. Finally we study the effect of particle losses on the squeezing of a nonlinear or a quadratic spin observable.

\section{Optimization over rotation axis and measurement observables}\label{sec:methode}
We consider an ensemble of $N$ two-level atoms that is described in terms of the collective spin observables $\hat{\vec{S}}=(\hat{S}_x,\hat{S}_y,\hat{S}_z)^T$, where $\hat{S}_{k}=\sum_{i=1}^N\hat{\sigma}^{(i)}_k/2$ and $\sigma^{(i)}_k$ with $k=x,y,z$ is the Pauli matrix for the $i$-th atom. Starting from the spin-coherent state $|\psi_0\rangle$ such that
\begin{align}\label{eq:CCS}
\hat{S}_x|\psi_0\rangle=\frac{N}{2}|\psi_0\rangle,
\end{align}
an entangled spin state $|\psi_t\rangle=\hat{U}_t|\psi_0\rangle$ is generated via the OAT evolution $\hat{U}_t=e^{-i\chi t \hat{S}_z^2}$  at time $t$. 
A phase $\phi$ is imprinted at this time by the rotation $e^{-i\hat{S}_{\vec{n}}\,\phi}$, with $\hat{S}_{\vec{n}}=\vec{n}\cdot\hat{\vec{S}}$, where $\vec{n}$ is a unit vector in the plane perpendicular to the initial spin polarization, $\vec{e}_x$ in this case. The goal of the protocol is to infer the best estimate of the phase $\phi$ from the measurement of an observable $\hat{X}$, subsequently to the phase imprinting.
The inferred phase uncertainty is given by \cite{WinelandPRA1992}
\begin{align}\label{eq:uncertainty}
(\Delta\phi)^2=\left.\frac{(\Delta \hat{X})^2}{|\partial_{\phi}\langle\hat{X}\rangle|^2}\right|_{\phi=0},
\end{align}
where $\partial_{\phi}\equiv \partial/\partial\phi$, while $\langle\hat{X}\rangle$ and $(\Delta \hat{X})^2$ are the mean value and the variance of the measured observable $\hat{X}$ respectively. Since any additional shift can be absorbed by the initial state, we focus on the estimation of the phase in the vicinity of zero without restriction of generality. The denominator in (\ref{eq:uncertainty}) is given by
\begin{equation}
\left.\frac{\partial \langle\hat{X}\rangle}{\partial\phi}\right|_{\phi=0}=\frac{\partial}{\partial\phi}\langle\psi_0|\hat{U}^{\dagger}_t e^{i\hat{S}_{\vec{n}}\,\phi}\hat{X}e^{-i\hat{S}_{\vec{n}}\,\phi}\hat{U}_t|\psi_0\rangle|_{\phi=0}
=i\langle\psi_t|[\hat{S}_{\vec{n}},\hat{X}]|\psi_t\rangle.
\end{equation}
By replacing it in (\ref{eq:uncertainty}), we obtain
\begin{equation}
(\Delta\phi)^2=\frac{(\Delta \hat{X})^2}{|\langle[\hat{S}_{\vec{n}},\hat{X}]\rangle|^2}.
\end{equation}
For the initial non-correlated state (\ref{eq:CCS}) and $\hat{X}=\hat{S}_{\vec{m}}$ a spin component in the $yz$-plane with $\vec{m}\perp\vec{n}$, the phase uncertainty reaches the SQL. With respect to this limit, we quantify the quantum metrological gain given by the state prepared at the time $t$ of the OAT evolution, with a rotation around $\vec{n}$ and a measurement of an observable $\hat{X}$, by the parameter~\cite{GessnerPRL2019}
\begin{align}\label{eq:Xidef}
\xi^{-2}(\chi t,\hat{S}_{\vec{n}},\hat{X})\equiv\frac{(\Delta\phi)_{\mathrm{SQL}}^2}{(\Delta\phi)^2}=\frac{|\langle[\hat{S}_{\vec{n}},\hat{X}]\rangle|^2}{N(\Delta\hat{X})^2},
\end{align}
where all the averages are taken in the state $\hat{U}_t|\psi_0\rangle$.
In order to analytically optimize the metrological gain (\ref{eq:Xidef}) with respect to the rotation axis $\vec{n}$ and the measurement observable $\hat{X}$, we assume that we have a family of $q$ accessible operators $\hat{\vec{X}}=(\hat{X}_1,...,\hat{X}_q)^T$ and we can measure any arbitrary linear combination $\hat{X}_{\vec{m}}=\vec{m}\cdot\hat{\vec{X}}= \sum_{k=1}^q m_k\hat{X}_k$. For a given measurement direction $\vec{m}$, we can re-express (\ref{eq:Xidef}) as
\begin{align}\label{eq:xireex}
\xi^{-2}(\chi t,\hat{S}_{\vec{n}},\hat{X}_{\vec{m}})=\frac{|\vec{n}^T C[\chi t,\hat{\vec{S}},\hat{\vec{X}}] \vec{m}|^2}{N (\vec{m}^T \Gamma[\chi t,\hat{\vec{X}}] \vec{m})}.
\end{align}
where we introduced the $2\times q$ commutator matrix
\begin{align}\label{eq:Cdef}
C[\chi t,\hat{\vec{S}},\hat{\vec{X}}]_{kl}\equiv-i\langle[\hat{S}_k,\hat{X}_l]\rangle.
\end{align}
and the $q\times q$ covariance matrix
\begin{equation}\label{eq:gammadef}
\Gamma[\chi t,\hat{\vec{X}}]_{kl}\equiv\mathrm{Cov}(\hat{X}_k,\hat{X}_l)
=\frac{1}{2}\langle \hat{X}_k \hat{X}_l+\hat{X}_l\hat{X}_k\rangle-\langle \hat{X}_k\rangle\langle \hat{X}_l\rangle.
\end{equation}
For a state prepared at time $t$ of the OAT evolution, the maximum of~(\ref{eq:xireex}) over the rotation direction $\vec{n}$ and the measurement direction $\vec{m}$ corresponds to the maximum eigenvalue $\lambda_{\rm max}$ of the matrix $C \Gamma^{-1}C^T$~\cite{GessnerPRL2019}:
\begin{align}\label{eq:Xilambda}
\xi^{-2}_{\hat{\vec{X}}}(\chi t)=\max_{\vec{m},\vec{n}}\xi^{-2}(\chi t,\hat{S}_{\vec{n}},\hat{X}_{\vec{m}})=\frac{\lambda_{\max}(C \Gamma^{-1}C^T)}{N},
\end{align}
and is reached with the choice $\vec{n}=\vec{n}_{\max}$ where $\vec{n}_{\max}$ is the eigenvector of $C \Gamma^{-1}C^T$ corresponding to $\lambda_{\max}$. The optimal measurement direction is $\vec{m}_{\mathrm{opt}}=\alpha\Gamma^{-1}C^T\vec{n}_{\max}$, where $\alpha \in \mathbb{R}$ is a normalization constant. 


\begin{figure*}[tb]
    \centering
    \includegraphics[width=\textwidth]{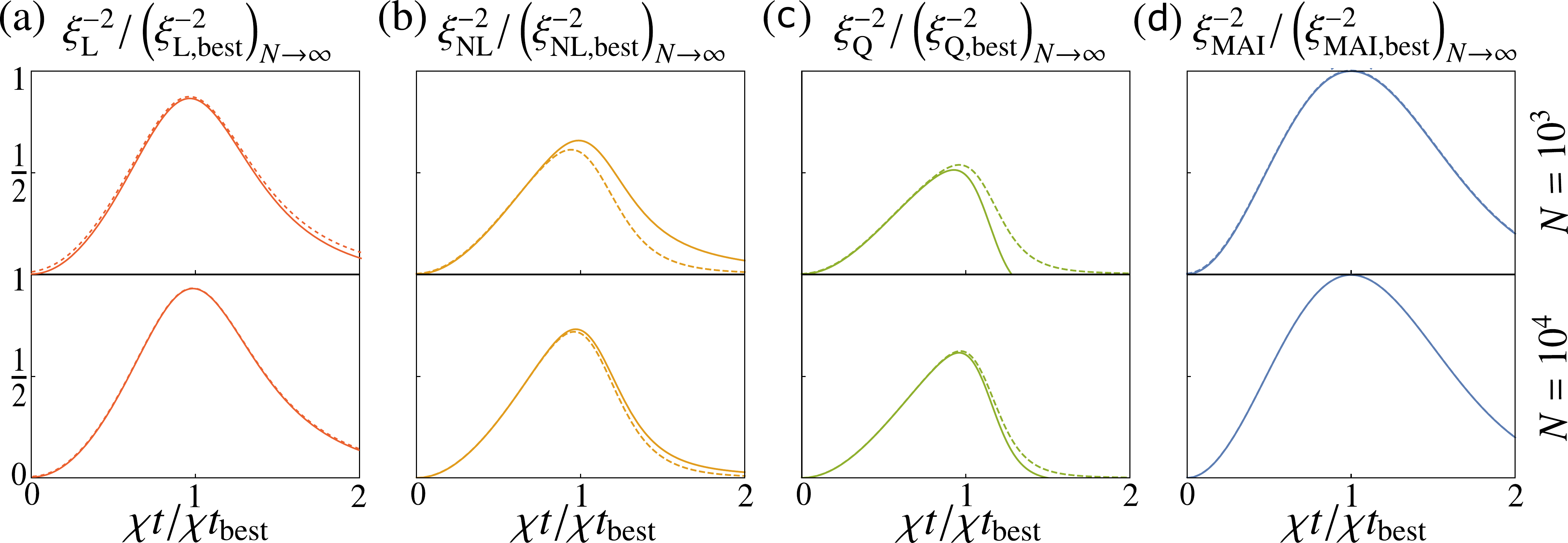}
    \caption{{ Metrological gain $\xi^{-2}$ in the limit of large $N$ including finite size corrections (\ref{eq:xifiniteL}), (\ref{eq:xifiniteNL}), (\ref{eq:xifiniteq}) and (\ref{eq:MAIxinorm}) (solid lines) for (a) linear (L), (b) nonlinear (NL), (c) quadratic (Q), and (d) MAI measurement strategies, compared to the exact metrological gain (dashed lines). The spin number is $N=10^3$ (top row) and $N=10^4$ (bottom row).}}
    \label{unified}
\end{figure*}

\section{Squeezing of linear and quadratic spin observables}
Starting with a coherent spin state, at short times, the OAT evolution $\hat{U}_t$ leads to the squeezing of a linear spin component $\hat{X}_{\rm L}=\hat{S}_{\vec{m}}$. 
An evolution beyond the best linear squeezing time allows for the generation of non-Gaussian spin states where nonlinear spin observables are squeezed. For each given choice of a family $\hat{\vec{X}}$ of accessible operators that can contain nonlinear spin observables, in addition to $\hat{S}_{\vec{m}}$, the optimization explained in Sec. \ref{sec:methode} allows us to identify, at any time $t$ of the one-axis-twisting evolution, the best squeezed observable and the corresponding metrological gain.
\subsection{Linear spin squeezing}\label{seq. FamiliesL}
Let us first consider the squeezing of a linear spin observable $\hat{X}_{\rm L}=\hat{S}_{\vec{m}}=\sum_{i=x,y,z}m_i\hat{S}_i$. By considering that the initial state of the system is (\ref{eq:CCS}) where the collective spin is in the $x$ direction, we can show that $\langle\psi_0|\hat{U}_t^{\dagger}[\hat{S}_x,\hat{S}_i]\hat{U}_t|\psi_0\rangle=0$ for any $i=x,y,z$. This allows us to restrict $\vec{m}$ to the $yz$-plane. In order to identify the best squeezed linear observable and the corresponding metrological gain, we use the technique explained in Sec. \ref{sec:methode} and we set $\hat{\vec{X}}_{\mathrm{L}}=\left(\hat{S}_y,\hat{S}_z\right)^T$, meaning that we study the squeezing of a linear observable of the form
\begin{align}\label{eq:Xl}
\hat{X}_{\mathrm{L}}=m_y\hat{S}_y+m_z\hat{S}_z.
\end{align}
The fact that the one-axis-twisting evolution is analytically solvable allows us to determine, see \ref{A}, the commutator (\ref{eq:Cdef}) and the covariance (\ref{eq:gammadef}) matrices for a given $N$ at each time $t$. The optimization over the rotation $\vec{n}$ and the measurement $\vec{m}$ directions gives us the metrological gain (\ref{eq:Xilambda}) in the limit $N\gg1$ at $\chi t< 1/\sqrt{N}$
\begin{equation}\label{eq:XiThLimitL}
\left(\xi^{-2}_{\rm L}(\chi t)\right)_{N\to\infty}=\frac{N^2(\chi t)^2}{1+N^4(\chi t)^{6}/6}.
\end{equation}
The best metrological gain and the corresponding time can be obtained from a maximization of (\ref{eq:XiThLimitL}) over $\chi t$ as \cite{SinatraFro2012}
\begin{equation}
\chi t_{\mathrm{L,best}}=3^{1/6} N^{-2/3} \qquad  ; \qquad  \left(\xi_{\mathrm{L,best}}^{-2}\right)_{N\rightarrow\infty}=\frac{2}{3^{2/3}} N^{2/3}.
\label{eq:bestL}
\end{equation}
By introducing the rescaled time $\tilde{\chi t}=\chi t/(\chi t_{\rm L,best})$ and by expanding the exact metrological gain $\xi^{-2}_{\rm L}(\chi t)$ up to $\mathcal{O}(N^{0})$, we obtain
\begin{equation}\label{eq:xifiniteL}
\quad\frac{\xi^{-2}_{\mathrm{L}}}{\left(\xi^{-2}_{\mathrm{L,best}}\right)_{N\to\infty}}=
\frac{3}{2}\frac{(\tilde{\chi t})^2}{1+(\tilde{\chi t})^{6}/2}
\left[1-3^{1/3}(\tilde{\chi t})^2 N^{-1/3}+\mathcal{O}(N^{-2/3})\right].
\end{equation}
This { expression is shown as a solid line in Fig.~\ref{unified}(a) as a function of $\tilde{\chi t}$ for $N=10^3,10^4$, and compared to the exact metrological gain. For $\tilde{\chi t}=1$ we obtain the best metrological gain including finite size corrections}
\begin{eqnarray}
\xi^{-2}_{\mathrm{L,best}}&=\left(\xi_{\mathrm{L,best}}^{-2}\right)_{N\rightarrow\infty}\left[1-3^{1/3} N^{-1/3}+\mathcal{O}(N^{-2/3})\right].\label{eq:XibestL}
\end{eqnarray}
that is shown as the red horizontal dashed line in Fig.~\ref{best_scaling}(a). 
The optimal rotation is $\hat{S}_{\vec{n}_{\max}}=\vec{n}_{\rm max}\cdot\hat{\vec{S}}_\perp$ where $\vec{n}_{\max}$ is a unit vector in the $yz$-plane with $\vec{n}_{\max}=(\cos\theta_{n}^{\rm L},\sin\theta_{n}^{\rm L})^T$, and the best squeezed linear spin observable is $\hat{S}_{\vec{m}_{\mathrm{opt}}}=\vec{m}_{\mathrm{opt}}\cdot\hat{\vec{X}}_{\mathrm{L}}$ with $\vec{m}_{\rm opt}=(\cos\theta_{m}^{\rm L},\sin\theta_{m}^{\rm L})^T$. In the limit of large $N$, we obtain
\begin{equation}
\theta_{n}^{\rm L}=3^{-1/6}N^{-1/3}+\mathcal{O}(N^{-2/3}) \qquad ; \qquad
\theta_{m}^{\rm L}=-\frac{\pi}{2}+3^{-1/6}N^{-1/3}+\mathcal{O}(N^{-2/3}).
\end{equation}
The interferometric estimation of the unknown phase $\phi$ using the state prepared at $\chi t_{\rm L, best}$ of the OAT dynamics with the rotation generator $\hat{S}_{\vec{n}_{\max}}$ and the measurement of the best squeezed linear observable $\hat{S}_{\vec{m}_{\rm opt}}$ lead to the sub-SQL phase uncertainty 
\begin{align}\label{eq:uncl}
\Delta\phi\simeq\frac{3^{1/3}}{\sqrt{2}}\frac{1}{N^{5/6}}.
\end{align}
\subsection{Nonlinear spin squeezing}\label{seq. Families}
In addition to $\hat{S}_{\vec{m}}$, we first consider a single second-order observable $\frac{1}{2}\{\hat{S}_x,\hat{S}_z\}$, where $\{\hat{A},\hat{B}\}=\hat{A}\hat{B}+\hat{B}\hat{A}$ denotes the anticomutator of $\hat{A}$ and $\hat{B}$. This corresponds to the choice of the nonlinear family $\hat{\vec{X}}_{\mathrm{NL}}=\left(\hat{S}_y,\hat{S}_z,\frac{1}{2}\{\hat{S}_x,\hat{S}_z\}\right)^T$. We thus explore the squeezing of a nonlinear observable of the form
\begin{align}\label{eq:Xnl}
\hat{X}_{\mathrm{NL}}=m_y\hat{S}_y+m_z\hat{S}_z+\frac{m_{xz}}{2}\{\hat{S}_x,\hat{S}_z\}.
\end{align}
The analytical calculation of the commutator (\ref{eq:Cdef}) and covariance (\ref{eq:gammadef}) matrices (given in \ref{A}), allows us to deduce the nonlinear metrological gain for $N\gg1$ at $\chi t< 1/\sqrt{N}$ as{\footnote{We calculate analytically the inverse of the $3\times 3$ covariance matrix $\Gamma$ and diagonalize the $2\times 2$ matrix $C\Gamma^{-1}C^T$.}}
\begin{equation}\label{eq:XiThLimit}
\left(\xi^{-2}_{\rm NL}(\chi t)\right)_{N\to\infty}=\frac{N^2(\chi t)^2}{1+N^6(\chi t)^{10}/270}.
\end{equation}
By maximizing (\ref{eq:XiThLimit}) over $\chi t$, { we find the scaling with $N$ of the} best metrological gain and the corresponding time:
\begin{equation}\label{eq:bestinf}
\chi t_{\mathrm{NL,best}}=\left(\frac{5}{2}\right)^{1/10} 3^{3/10} N^{-3/5} \qquad ; \qquad \left(\xi_{\mathrm{NL,best}}^{-2}\right)_{N\rightarrow\infty}=2\left(\frac{2}{5}\right)^{4/5} 3^{3/5} N^{4/5}.
\end{equation}
In order to obtain the first finite-size corrections to (\ref{eq:XiThLimit}), we introduce the rescaled time 
$\tilde{\chi t}=\frac{\chi t}{\chi t_{\rm NL,best}}$ to obtain
\begin{equation}
\frac{\xi^{-2}_{\mathrm{NL}}}{\left(\xi^{-2}_{\mathrm{NL,best}}\right)_{N\to\infty}} = \frac{5(\tilde{\chi t})^2}{4+(\tilde{\chi t})^{10}}\left[1-\left(\frac{135}{2}\right)^{1/5}(\tilde{\chi t})^2N^{-1/5}+
\frac{5(\tilde{\chi t})^4}{4+(\tilde{\chi t})^{10}}\: \frac{221 (\tilde{\chi t}) ^{10}+672}
{40500^{1/5} \,28}\: N^{-2/5}+\mathcal{O}(N^{-3/5})\right].
\label{eq:xifiniteNL}
\end{equation}
A representation of (\ref{eq:xifiniteNL}) as a function of $\tilde{\chi t}$ for $N=10^3,10^4$ compared to the exact metrological gain is shown in Fig.~\ref{unified}(b). For $\tilde{\chi t}=1$, we obtain { the best nonlinear metrological gain including finite-size corrections}
\begin{equation}
\xi^{-2}_{\mathrm{NL,best}}=\left(\xi_{\mathrm{NL,best}}^{-2}\right)_{N\rightarrow\infty}\left[1-\left(\frac{5}{2}\right)^{1/5} 3^{3/5} N^{-1/5}\right. 
\left.+\frac{893}{2^{2/5}3^{4/5}5^{3/5}28}N^{-2/5}+\mathcal{O}(N^{-3/5})\right].
\label{eq:XibestNL}
\end{equation}
shown as the orange horizontal dashed line in Fig.~\ref{best_scaling}(a). 

The optimal rotation direction $\vec{n}_{\max}=(\cos\theta_{n}^{\rm NL},\sin\theta_{n}^{\rm NL})^T$ is given in the limit of large $N$ by
\begin{align}
\theta_{n}^{\rm NL}=\left(\frac{2}{5}\right)^{1/10}3^{-3/10}N^{-2/5}+\mathcal{O}(N^{-3/5})\,,
\end{align}
{ and the best spin observable among the nonlinear family $\vec{X}_{\mathrm{NL}}$ is $\hat{X}_{\vec{m}_{\mathrm{opt}}}=\vec{m}_{\mathrm{opt}}\cdot\hat{\vec{X}}_{\mathrm{NL}}$ where we write
\begin{equation}
\vec{m}_{\rm opt}=(\sin\varphi_{m}^{\rm NL}\cos\theta_{m}^{\rm NL},\sin\varphi_{m}^{\rm NL}\sin\theta_{m}^{\rm NL},\cos\varphi_{m}^{\rm NL})^T \qquad  \mbox{with} \quad
\theta_{m}^{\rm NL} \in [0,2\pi] \quad \mbox{and} \quad \varphi_{m}^{\rm NL}\in [0,\pi]
\end{equation}
and, in the limit of large $N$ we find,
\begin{equation}
\theta_{m}^{\rm NL}=-\frac{\pi}{2}+\frac{3^{7/10}}{2^{9/10}5^{1/10}}N^{-2/5}+\mathcal{O}(N^{-3/5}) \qquad ; \qquad
\varphi_{m}^{\rm NL}=\frac{\pi}{2}+\frac{1}{N}+\mathcal{O}(N^{-6/5}).
\label{eq:nlcomp}
\end{equation}
{ Note that, since $\hat{S}_x$ is of order of $N$, the contribution $m_{xz}\{\hat{S}_x,\hat{S}_z\}$ of the nonlinear observable to $\hat{X}_{\rm NL}$ (\ref{eq:Xnl}) is comparable to that of the linear observable although $m_{xz}=\cos\varphi_{m}^{\rm NL}$ is of order $1/N$}. 
{ If a phase $\phi$ is imprinted in the system at $\chi t_{\rm NL,best}$ after the OAT evolution, the measurement of $\hat{X}_{\rm NL}$ allows us to estimate the value of the phase $\phi$ with an uncertainty }
%
\begin{align}\label{eq:uncnl}
\Delta\phi\simeq\frac{1}{\sqrt{2}}\left(\frac{5}{2}\right)^{4/10}3^{-3/10}\frac{1}{N^{9/10}}\,,
\end{align}
clearly surpassing the squeezing of a linear observable (\ref{eq:uncl}) and approaching the Heisenberg limit $\Delta\phi= 1/N$.}
\subsection{Quadratic spin squeezing}
We now explore the squeezing of an arbitrary linear combination of spin observables up to second order. First, we find numerically that in the time window $0<\chi t\leq1/\sqrt{N}$ of the one-axis-twisting evolution, the best squeezed quadratic observable is a combinaition of only four observables $\hat{\vec{X}}_{\rm Q}=(\hat{S}_y,\hat{S}_z,\frac{1}{2}\{\hat{S}_x,\hat{S}_z\},\frac{1}{2}\{\hat{S}_x,\hat{S}_y\})^T$. For this reason, we limit ourselves, in the following, to the observables $\hat{X}_{\mathrm{Q}}$ of the form
\begin{align}\label{eq:Xnlq}
\hat{X}_{\mathrm{Q}}=m_y \hat{S}_y+m_z \hat{S}_z +\frac{m_{xz}}{2} \{\hat{S}_x,\hat{S}_z\}+\frac{m_{xy}}{2} \{\hat{S}_x,\hat{S}_y\}.
\end{align}
{ By proceeding similarly to the nonlinear case\footnote{
We calculate this time the inverse of the $4\times 4$ covariance matrix which is still analytically possible, and we take the limit $N\rightarrow\infty$ in the metrological gain $\xi^{-2}_{\rm Q}(\chi t)$ calculated by (\ref{eq:Xilambda}). The elements of the covariance and commutator matrices for the quadratic case are given in \ref{A}.}} we obtain for $\chi t<1/\sqrt{N}$: 
\begin{figure}[tb]
    \centering
    \includegraphics[width=0.9\textwidth]{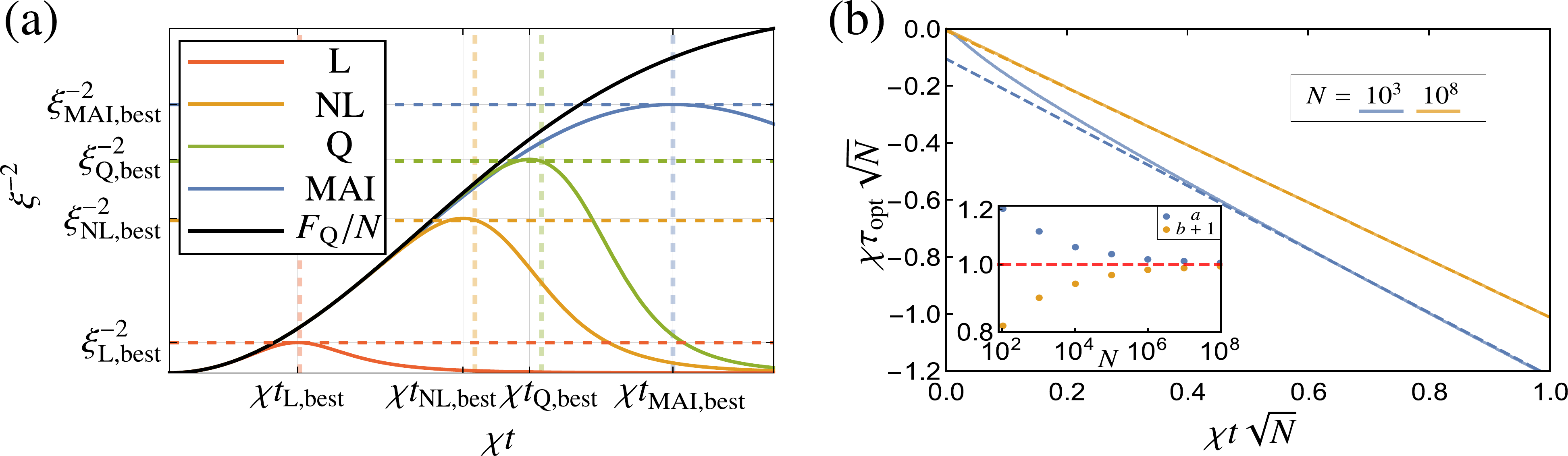}
    \caption{(a) Quantum metrological gain { for the linear $\xi^{-2}_{\mathrm{L}}$, nonlinear $\xi^{-2}_{\mathrm{NL}}$, quadratic $\xi^{-2}_{\mathrm{Q}}$  and MAI $\xi^{-2}_{\mathrm{MAI}}$ measurement strategies as a function of time, compared to the quantum Fisher information with $N=10^4$. The solid vertical and horizontal lines represent the corresponding (exact) best meteorological gain and best time, while the dashed horizontal and vertical lines represent the analytical scaling laws in the limit of large $N$ with finite size corrections}. (b) Optimisation over the second interaction time $\tau$ { in the MAI technique. The plot shows the best time $\tau_{\mathrm{opt}}$} as a function of the squeezing time $t$ for $N=10^{3}$ (blue) and $N=10^8$ (orange). In the relevant time frame $\chi t\leq 1/\sqrt{N}$, we have $\tau_{\mathrm{opt}}\simeq -a t +b$ where $a$ and $b+1$ are represented in the inset as functions of $N$.}
    \label{best_scaling}
\end{figure}

\begin{equation}\label{eq:XiThLimitq}
\left(\xi^{-2}_{\rm Q}(\chi t)\right)_{N\to\infty}=\frac{N^2(\chi t)^2}{1+N^8(\chi t)^{14}/875}.
\end{equation}
{ The best metrological gain and the corresponding time are obtained by maximizing (\ref{eq:XiThLimitq}) over $\chi t$:}
\begin{equation}
\chi t_{\mathrm{Q,best}}\simeq\left(\frac{7}{6}\right)^{1/14} 5^{3/14} N^{-4/7} \qquad ; \qquad \left(\xi_{\mathrm{Q,best}}^{-2}\right)_{N\rightarrow\infty}=\left(\frac{6}{7}\right)^{6/7} 5^{3/7} N^{6/7}.
\label{eq:bestxiq}
\end{equation}
Introducing the rescaled time $\tilde{\chi t}=\frac{\chi t}{\chi t_{\rm Q,best}}$ we obtain the first finite-size corrections to (\ref{eq:XiThLimitq}) as
\begin{align}\label{eq:xifiniteq}
\quad\frac{\xi^{-2}_{\mathrm{Q}}}{\left(\xi^{-2}_{\rm Q,best}\right)_{N\to\infty}}=\frac{7(\tilde{\chi t})^2}{6+(\tilde{\chi t})^{14}}
\left[1-\left(\frac{875}{6}\right)^{1/7}(\tilde{\chi t})^2  N^{-1/7} + \frac{7(\tilde{\chi t})^4}{6+(\tilde{\chi t})^{14}}
\: \frac{86 (\tilde{\chi t})^{14}+297}{5^{1/7}6^{2/7}7^{5/7}15}\: N^{-2/7} \notag \right. \\ \left. \quad
-\frac{7(\tilde{\chi t})^6}{6+(\tilde{\chi t})^{14}} \: \frac{61(\tilde{\chi t})^{14}+147}{3^{10/7}5^{5/7}7^{4/7}2^{3/7}}\: N^{-3/7}
+\mathcal{O}(N^{-4/7})\right]
\end{align}
represented in Fig.~\ref{unified}(c), and finite size corrections to the best quadratic metrological gain $\xi^{-2}_{\rm Q,best}$
\begin{equation}\label{eq:XibestQ}
\xi^{-2}_{\mathrm{Q,best}}=\left(\xi_{\mathrm{Q,best}}^{-2}\right)_{N\rightarrow\infty}\left[1-\left(\frac{7}{6}\right)^{1/7} 5^{3/7} N^{-1/7}+\frac{383}{5^{1/7}6^{2/7}7^{5/7}15}N^{-2/7}
-\frac{104\times 2^{4/7}}{3^{10/7}5^{5/7}7^{4/7}}N^{-3/7}+\mathcal{O}(N^{-4/7})\right]
\end{equation}
that is represented as the green dashed horizontal line in Fig.~\ref{best_scaling}(a). 

The optimal rotation direction is $\vec{n}_{\rm max}=(\cos\theta_{n}^{\rm Q},\sin\theta_{n}^{\rm Q})^T$, and 
the best observable is $\hat{X}_{\vec{m}_{\rm opt}}=\vec{m}_{\rm opt}\cdot\hat{\vec{X}}_{\rm Q}$ where $\vec{m}_{\rm opt}$ is in this case a four-dimensional unit vector corresponding to the set of observables $(\hat{S}_y,\hat{S}_z,\frac{1}{2}\{\hat{S}_x,\hat{S}_z\},\frac{1}{2}\{\hat{S}_x,\hat{S}_y\})^T$ that can be written as 
\begin{equation}
\vec{m}_{\rm opt}=(\sin\omega_m^{\rm Q}\sin\varphi_m^{\rm Q}\cos\theta_m^{\rm Q}, \sin\omega_m^{\rm Q}\sin\varphi_m^{\rm Q}\sin\theta_m^{\rm Q}, \sin\omega_m^{\rm Q}\cos\varphi_m^{\rm Q}, \cos\omega_m^{\rm Q})^T \,.
\end{equation}
In the limit $N\gg 1$, we obtain
\begin{eqnarray}
\theta_{n}^{\rm Q}&=&\left(\frac{6}{7}\right)^{1/14}5^{-3/14}N^{-3/7}+\mathcal{O}(N^{-4/7}) \qquad;\qquad
\theta_m^{\rm Q}=-\frac{\pi}{2}+\frac{2^{15/14}}{3\times 5^{3/14}}\left(\frac{7}{3}\right)^{13/14}N^{-3/7}+\mathcal{O}(N^{-4/7}) \\
\varphi_m^{\rm Q}&=&\frac{\pi}{2}-\frac{4}{3N}+\mathcal{O}(N^{-8/7}) \qquad ; \qquad \qquad \qquad
\omega_m^{\rm Q}=\frac{\pi}{2}+\frac{2}{3}\left(\frac{2}{7}\right)^{1/14}\frac{1}{3^{13/14}5^{3/14}}N^{-10/7}+\mathcal{O}(N^{-11/7}).
\end{eqnarray}
{ By taking into the account that $\hat{S}_x$ is of the order of $N$, we note that the contribution of the two nonlinear observables $\{\hat{S}_x,\hat{S}_z\}$ and $\{\hat{S}_x,\hat{S}_y\}$ to $\hat{X}_{\vec{m}_{\rm opt}}$ are respectively of the same order and $N^{-3/7}$ smaller than the contribution of the linear observable}. The squeezing of the quadratic observable~(\ref{eq:Xnlq}) allows to achieve an { uncertainty
 \begin{align}\label{eq:uncq}
\Delta\phi\simeq\left(\frac{7}{6}\right)^{6/14} 5^{-3/14} \frac{1}{N^{13/14}},
\end{align}
on the inferred phase} which is even closer to the Heisenberg limit than the uncertainty~(\ref{eq:uncnl}) attained by the squeezing of the nonlinear observable~(\ref{eq:Xnl}). As expected, the uncertainty on the phase $\Delta\phi$ decreases as { both the preparation time of the state by OAT evolution and the nonlinearity of the measured spin observable increase}.

\section{Scaling laws of measurement-after-interaction technique}
\label{sec:MAI}
As shown above, { the evolution with the one-axis twisting Hamiltonian, used as a system preparation before phase imprinting,} allows to achieve a high metrological gain through the squeezing of nonlinear spin observables. Such observables, that are higher moments of the spin components, can be extracted from the statistics of linear spin observables~\cite{LuckeSCIENCE2011,StrobelSCIENCE2014,BohnetSCIENCE2016,EvrardPRL2019,XuArXiv2021}. However, due to the increased measurement time and the need for low detection noise, this is challenging to achieve in systems with large atom numbers. As we will show in this section, the MAI technique~\cite{SchleierSmithPRL2016,FrowisPRL2016,NolanPRL2017} represents an alternative method for measuring a nonlinear spin observable directly. For that, after the phase impinting and prior to the measurement of a linear spin observable $\hat{S}_{\vec{m}}$ with $\vec{m}=(m_x,m_y,m_z)^T$, we allow a second evolution $\hat{U}_{\tau}$ of the system with the OAT Hamiltonian $\hat{H}=\hbar\chi\hat{S}_z^2$. Mathematically, this is equivalent to the measurement of the nonlinear spin observable 
\begin{align}\label{eq:maix}
\hat{X}_{\rm MAI}=\hat{U}_{\tau}^{\dagger}\hat{S}_{\vec{m}}\hat{U}_{\tau}=\sum_{k=x,y,z}m_k e^{i\chi\tau\hat{S}_z^2}\hat{S}_k e^{-i\chi\tau\hat{S}_z^2}.
\end{align}
By expanding (\ref{eq:maix}) up to linear order in $\chi \tau$, we obtain
\begin{align}\label{eq:maixapp}
\hat{X}_{\rm MAI}=\hat{S}_{\vec{m}}{ -\chi\tau m_x\{\hat{S}_y,\hat{S}_z\}}+\chi\tau m_y\{\hat{S}_x,\hat{S}_z\}+ \mathcal{O}(\chi\tau)^2.
\end{align}
Hence, a OAT evolution up to $\chi\tau= m_{xz}/(2m_y)$ followed by a measurement of the linear spin observable $\hat{S}_{\vec{m}}$ { with $m_x=0$} is equivalent to first order in $\chi\tau$ to the measurement of the
nonlinear spin observable~(\ref{eq:Xnl}). 

Motivated by this correspondence, we systematically study the metrological potential that is offered by the continuous set of observables~(\ref{eq:maix}), which is parametrized by $\chi \tau$ and accessible by the MAI technique. The analytical optimization (\ref{eq:Xilambda}) allows us to obtain the maximal metrological gain $\xi^{-2}_{\mathrm{MAI}}(\chi t)$ over all rotation directions $\vec{n}$ and measurement directions $\vec{m}$ for a fixed interaction time $\tau$ at time $t$ of the OAT evolution{ \footnote{{ Starting with the coherent spin state (\ref{eq:CCS}), the metrological gain associated to the state prepared at time $\chi t$ of the OAT evolution with the measurement of the observable~(\ref{eq:maix}) for fixed $\chi\tau$ is written according to~(\ref{eq:xireex}), (\ref{eq:Cdef}) and (\ref{eq:gammadef}), with $\vec{X}=\hat{U}_{\tau}^{\dagger}\hat{\vec{S}}\hat{U}_{\tau}$. We thus have to evaluate
\begin{align}\label{c}
C[\chi t,\hat{\vec{S}},\hat{U}_{\tau}^{\dagger}\hat{\vec{S}}\hat{U}_{\tau}]_{kl}=-i\langle[\hat{S}_k,\hat{U}_{\tau}^{\dagger}\hat{S}_l \hat{U}_{\tau}]\rangle,
\qquad \mbox{and} \qquad
\Gamma[\chi t,\hat{U}_{\tau}^{\dagger}\hat{\vec{S}}\hat{U}_{\tau}]_{kl}=\mathrm{Cov}(\hat{U}_{\tau}^{\dagger}\hat{S}_k \hat{U}_{\tau},\hat{U}_{\tau}^{\dagger}\hat{S}_l \hat{U}_{\tau})=\Gamma[\chi (t+\tau),\hat{\vec{S}}]_{kl},
\end{align}
where we used the property $\hat{U}_t\hat{U}_{\tau}=\hat{U}_{t+\tau}$. First, we note that $\langle\psi_0|\hat{U}^{\dagger}_t[\hat{S}_x,\hat{U}^{\dagger}_{\tau}\hat{S}_{\vec{m}}\hat{U}_{\tau}]\hat{U}_t|\psi_0\rangle=\langle\psi_0|\hat{U}^{\dagger}_t[\hat{S}_{\vec{n}},\hat{U}^{\dagger}_{\tau}\hat{S}_{x}\hat{U}_{\tau}]\hat{U}_t|\psi_0\rangle=0$ for any linear spin observable $\hat{S}_{\vec{m}}$ and $\hat{S}_{\vec{n}}$. This allows us to restrict the optimization of both the rotation direction $\vec{n}$ and the measurement direction $\vec{m}$ to the plane perpendicular to the initial spin direction $\vec{e}_x$. The $2\times 2$ commutator and covariance matrices are given in \ref{B}.}}}. 

As shown in Fig.~\ref{best_scaling}(b), numerical optimization over $\tau$ reveals that, in the limit of large $N$, for a given $\chi t\leq 1/\sqrt{N}$, the optimal interaction time $\chi\tau_{\mathrm{opt}}$ which maximizes the metrological gain $\xi^{-2}_{\rm MAI}(\chi t)$ is given by
\begin{align}\label{eq:tauopt}
\chi\tau_{\mathrm{opt}}\underset{N\gg 1}{\to}-\chi t.
\end{align}
This corresponds to the echo protocol that was first suggested in Ref.~\cite{SchleierSmithPRL2016} where, after the first one-axis-twisting evolution up to $t$ and phase imprinting, we implement a second one-axis-twisting evolution of a duration $t$ where we invert the sign of the constant $\chi\to -\chi$ in the nonlinear Hamiltonian. Motivated by the result (\ref{eq:tauopt}), we replace $\chi\tau$ by $-\chi t$ in the expression of the observable $\hat{X}_{\rm MAI}$ (\ref{eq:maix}). Using $\cos^N(\chi t)\simeq e^{-N(\chi t)^2/2}$ for $\chi t\to 0$, the metrological gain $\xi_{\rm MAI}^{-2}(\chi t)$ for the MAI technique is given for $\chi t\leq 1/\sqrt{N}$ in the limit of large $N$ by 
\begin{align}\label{eq:MAIxi}
\left(\xi_{\rm MAI}^{-2}(\chi t)\right)_{N\to\infty}=N^2(\chi t)^2e^{-N(\chi t)^2}.
\end{align}
The scaling laws of the metrological gain $\xi_{\rm MAI}^{-2}$ for $N\gg 1$ on the time scales
\begin{align}\label{eq:tscal}
\chi t=\sigma N^{-\alpha}, &\quad 1\geq\alpha \geq 1/2,
\end{align}
can easily obtained from Eq.~(\ref{eq:MAIxi}) and read\footnote{For $1\geq\alpha > 1/2$, we do not include the first correction whose form depends on the value of $\alpha$}
\begin{align}\label{eq:MAIopt}
\xi^{-2}_{\mathrm{MAI}}=
\begin{cases}
\sigma^2 N^{2-2\alpha}, &\: 1\geq\alpha > 1/2\\
&\\
\sigma^2e^{-\sigma^2} N\left[1+(\frac{1+e^{\sigma^2}}{\sigma^2}+\frac{5\sigma^2}{3}
-\frac{\sigma^4}{6}-2)\frac{1}{N}+\mathcal{O}(N^{-2})\right], &\:  \alpha = 1/2
\end{cases}.
\end{align}
We first note that to the leading order in the limit of large $N$, the result (\ref{eq:MAIopt}) reproduces the scaling laws of the metrological gain of the linear, the nonlinear and the quadratic spin squeezing discussed above:  for $\alpha=2/3$, we recover the scaling of $\xi^{-2}_{\rm L}\propto N^{2/3}$ for the linear spin squeezing.  For $\alpha=3/5$, the scaling law $\xi^{-2}_{\rm NL}\propto N^{4/5}$ of the squeezing of the nonlinear observable (\ref{eq:Xnl}) and for $\alpha=4/7$, the scaling law $\xi^{-2}_{\rm Q}\propto N^{6/7}$ of the squeezing of a quadratic observable. The best metrological gain of the echo protocol~\cite{SchleierSmithPRL2016}, yielding the Heisenberg scaling $\xi^{-2}_{\mathrm{MAI,best}}=N/e$ at the time $\chi t_{\mathrm{MAI,best}}=1/\sqrt{N}$, is obtained from (\ref{eq:MAIopt}) by maximization over both $\sigma$ and $\alpha$. Simlarly to the previous section, the time rescaling $\tilde{\chi t}=\chi t/\chi t_{\rm MAI,best}$ allows us to write
\begin{align}\label{eq:MAIxinorm}
\frac{\xi^{-2}_{\rm MAI}}{\xi^{-2}_{\rm MAI,best}}=(\tilde{\chi t})^2e^{1-(\tilde{\chi t})^2}+\mathcal{O}(N^{-1}).
\end{align}
This is represented in Fig.~\ref{unified}(d). Note that the first finite size correction to the metrological gain (\ref{eq:MAIxinorm}) of the MAI method, { of order} $1/N$, are very small compared to the case of the
nonlinear ($1/N^{1/5}$) and the quadratic ($1/N^{1/7}$ ) spin squeezing. 

The optimal rotation direction for a given $\alpha$ and $\sigma$ is written as $\vec{n}_{\rm max}=\cos\theta_n\vec{e}_y+\sin\theta_{n}\vec{e}_z$ where we obtain in the limit of large $N$
\begin{align}\label{eq:MAIrot}
\theta_{n}=
\begin{cases}
\arctan (\frac{2}{\sigma}N^{\alpha-1}), &\: 1\geq\alpha > 1/2\\
\frac{1}{\sigma}e^{\sigma^2/2}\frac{1}{\sqrt{N}}+\mathcal{O}(N^{-3/2}),&\:  \alpha = 1/2
\end{cases}.
\end{align}
$\hat{X}_{\vec{m}_{\rm opt}}=\hat{U}_{-t}^{\dagger}(\vec{m}_{\rm opt}\cdot\hat{\vec{S}})\hat{U}_{-t}$, where $\vec{m}_{\rm opt}=\cos\theta_{m}\vec{e}_y+\sin\theta_{m}\vec{e}_z$ is a unit vector with
\begin{align}\label{eq:MAImes}
\theta_{m}=
\begin{cases}
\arctan (-\frac{1}{\sigma}N^{\alpha-1}), &\: 1\geq\alpha > 1/2\\
-\frac{1}{\sigma}\frac{1}{\sqrt{N}}+\mathcal{O}(N^{-1}),&\:  \alpha = 1/2
\end{cases},
\end{align}
represents, among the continuous set of observables (\ref{eq:maix}), the best squeezed nonlinear observable at the time (\ref{eq:tscal}) of the one-axis-twisting evolution\footnote{Here again in (\ref{eq:MAIrot}) and (\ref{eq:MAImes}), the first corrections for $1\geq\alpha > 1/2$ depend on the value of $\alpha$.}. For $\alpha=1/2$, Eqs.~(\ref{eq:MAIrot}) and~(\ref{eq:MAImes}) confirm the optimality of the rotation direction $\vec{n}=\vec{e}_y$ and the measurement direction $\vec{m}=\vec{e}_y$ made in Ref.~\cite{SchleierSmithPRL2016} for $N\to \infty$.

\section{Quantum Fisher information}\label{sec:QFI}
The full metrological potential of a state is given { by the quantum Fisher information $F_Q$~\cite{BraunsteinPRL1994} obtained by optimization over all possible measurements} $\max_{\hat{X}}\xi^{-2}=F_Q/N$. In order to assess the quality of the MAI technique, we compare $\xi^{-2}_{\rm MAI}$, given in Eq.~(\ref{eq:MAIopt}), to $F_Q/N$ of the { states generated by one-axis twisting}. Starting with the state~(\ref{eq:CCS}), for a phase imprinting rotation around $\hat{S}_{\vec{n}}$ with $\vec{n}$ in the $yz$-plane, the quantum Fisher information at a time $t$ of the one-axis-twisting evolution is given by $F_Q=4\lambda_{\rm max,F}$, where $\lambda_{\rm max,F}$ is the largest eigenvalue of the covariance matrix $\Gamma[\chi t,\hat{\vec{X}}_{yz}=(\hat{S}_y,\hat{S}_z)^T]$ \cite{PezzeRMP2018}. This can be obtained by restricting the covariance matrix of the quadratic measurement given in \ref{A} to the first two rows and columns. In the limit of large $N$ { and for} $\chi t\leq 1/\sqrt{N}$, we obtain
\begin{align}\label{eq:Fisher}
\left(F_Q/N\right)_{N\to\infty}=\frac{1}{2}\left(1-e^{-2 N (\chi t)^2}\right)N.
\end{align} 
Using (\ref{eq:Fisher}), we obtain the scaling law of the quantum Fisher information $F_Q/N$ at the time scales $\chi t=\sigma N^{-\alpha}$ with $1\geq\alpha\geq 1/2$
\begin{align}\label{eq:Fisheropt}
F_Q/N=
\begin{cases}
\sigma^2 N^{2-2\alpha}, & 1\geq\alpha > 1/2\\
\frac{1}{2}(1-e^{-2\sigma^2}) N+\mathcal{O}(N^{0}), &\: \alpha = 1/2
\end{cases}.
\end{align}
{ Comparison of this last equation to (\ref{eq:MAIopt}), shows} that the MAI technique reaches the optimal scaling law of sensitivity enhancement over the entire range of time $1\geq\alpha > 1/2$.

We note that the metrological gain $\xi^{-2}_{\rm L}$ (\ref{eq:XiThLimitL}), $\xi^{-2}_{\rm NL}$ (\ref{eq:XiThLimit}) and $\xi^{-2}_{\rm Q}$ (\ref{eq:XiThLimitq}) discussed above have the same structure and can be summarized in a unifying formula that gives the metrological gain in the limit of large $N$ for different mesurement strategies. For $\chi t<1/\sqrt{N}$, we have
\begin{equation}\label{eq:XiThLimitGen}
\left(\xi^{-2}(\chi t)\right)_{N\to\infty}=\frac{F_Q/N}{1+M},
\end{equation}
where 
\begin{equation}
M_{\rm L}=\frac{N^4(\chi t)^{6}}{6} \qquad ; \qquad
M_{\rm NL}=\frac{N^6(\chi t)^{10}}{270}  \qquad ; \qquad
M_{\rm Q}=\frac{N^8(\chi t)^{14}}{875}
\end{equation}
for a linear, nonlinear and quadratic measurement respectively. In the case of the MAI technique, the metrological gain in the limit of large $N$ is given by (\ref{eq:XiThLimitGen}) for $\chi t\leq1/\sqrt{N}$ with
\begin{align}
M_{\rm MAI}=\frac{\sinh[N(\chi t)^2]}{N(\chi t)^2}-1.
\end{align}
These expressions quantify the limitation of the metrological gain due to suboptimal measurements ($M$). In this sense, $M$ can be interpreted as the information that cannot be extracted from the state in a given measurement strategy.

\section{Dephasing noise}
In experiments, { for physical systems that are not perfectly isolated from the environment or that have other degrees of freedom coupled to the spin degrees of freedom we are interested in,} decoherence affects the OAT evolution and limits the metrological gain $\xi^{-2}$. Realizations of the OAT evolution based on Bose-Einstein condensates are { fundamentally} limited by particle losses and finite temperature~\cite{LiYunPRL2008,SinatraPRL2011}. It has been shown that for spin squeezing these effects can be described with a dephasing model that leads to a ballistic behavior of spin fluctuations $(\Delta\hat{S}_y)^2$~\cite{SinatraFro2012}. In OAT realizations using trapped ions~\cite{MolmerPRL1999,MonzPRL2011,BohnetSCIENCE2016}, magnetic field fluctuations cause a similar ballistic collective dephasing~\cite{MonzPRL2011,LanyonPRL2013,CarnioPRL2015}. On the contrary, in cavity-induced squeezing of atomic ensembles, the collective dephasing of the spin due to cavity losses is of a diffusive nature~\cite{LerouxPRA2012, PawlowskiEPL2016}. In the following, we focus on these classes of processes, i.e. on ballistic or diffusive fluctuations of a collective spin observable and we quantify the resulting limitations on the metrological gain $\xi^{-2}$. The ballistic dephasing model is based on a Hamiltonian evolution with a parameter that fluctuates from a realization to the other, which on average, leads to incoherent evolution. The diffusive dephasing model is obtained from a Lindblad master equation~\cite{Tannoudji,Breuer}.

\subsection{Ballistic dephasing}\label{Bal}
To describe the OAT evolution in the presence of a ballistic collective dephasing, we consider the Hamiltonian 
\begin{equation}\label{Hbal}
\hat{H}_{\mathrm{bal}}=\hbar \chi (\hat{S}_z^2 + D \hat{S}_z),
\end{equation}
where, $\chi D$ represents an energy shift in the two-level systems. The constant $D$, here, is a classical random variable whose value fluctuates between different repetitions of the experiment. We consider $D$ to follow a Gaussian distribution $p(D)$
{ with zero average and a possibly extensive variance
\begin{equation}
p(D)=\frac{1}{\sqrt{2\pi \langle D^2 \rangle}}e^{-\frac{D^2}{2\langle D^2 \rangle}} 
\qquad \mbox{where} \qquad \langle D^2 \rangle = \epsilon  N^\gamma \quad \textrm{with} \quad 0 \leq \gamma \leq 1,
\end{equation}
and $\epsilon$ a small parameter.} 
Starting again with the coherent spin state (\ref{eq:CCS}), the state of the system becomes  $|\psi_t\rangle=e^{-i\hat{H}_{\rm bal}t/\hbar}|\psi_0\rangle$, and the expectation value $\langle \hat{A}\rangle$ of any observable $\hat{A}$ is given by
\begin{align}
\langle \hat{A}\rangle=\int_{-\infty}^{+\infty}p(D)\langle\psi_t|\hat{A}|\psi_t\rangle dD.
\end{align}

\subsubsection{Linear, nonlinear, and quadratic spin observables}
The metrological gain of the state $|\psi_t\rangle$, with a rotation around $\vec{n}$ and a measurement of $\hat{X}$ can always be written as in Eq.~(\ref{eq:xireex}) where the corresponding analytical expressions for $C$ and $\Gamma$ are given in~\ref{C}. Following an analogous strategy as in the noiseless case, to the leading order in the limit of large $N$, the metrological gain of the linear, the nonlinear and the quadratic spin squeezing is obtained for $\chi t< 1/\sqrt{N}$ as
\begin{equation}\label{eq:bal}
\left(\xi^{-2}_{\rm bal}(\chi t)\right)_{N\to\infty}=\frac{N^2(\chi t)^2}{1+M+\epsilon N^{1+\gamma}(\chi t)^2},
\end{equation}
with the appropriate expression $M$ of each measurement strategy given in Sec.~\ref{sec:QFI}. The precise scaling in the large-$N$ limit now depends on the interplay between the terms in the denominator. Generally, we note that as soon as the noise-dependent term becomes non-negligible over $1+M$, it will determine the scaling of the maximal quantum gain. Thus, the effect of ballistic dephasing, in the limit of large $N$, is to set the upper bound $\xi^{-2}_{\lim}=N^{1-\gamma}/\epsilon$ to the scaling of the metrological gain, independently of the measurement strategy. { Due to the form of~(\ref{eq:bal}), the maximisation  over $\chi t$ is not affected by the ballistic dephasing. The best time $\chi t_{\rm best}$ is then unchanged and the} metrological gain is
\begin{equation}\label{eq:bal_all}
\xi_{\mathrm{best,L (bal)}}^{-2}\simeq\frac{2\times 3^{-2/3}N^{2/3}}{1+2\epsilon\times 3^{-2/3} N^{\gamma-1/3}}\quad ; \quad
\xi_{\mathrm{best,NL (bal)}}^{-2}\simeq\frac{2\left(\frac{2}{5}\right)^{4/5}3^{3/5}N^{4/5}}{1+2\epsilon\left(\frac{2}{5}\right)^{4/5}3^{3/5} N^{\gamma-1/5}}\quad ; \quad
\xi_{\mathrm{best,Q (bal)}}^{-2}\simeq\frac{\left(\frac{6}{7}\right)^{6/7}5^{3/7}N^{6/7}}{1+\epsilon\left(\frac{6}{7}\right)^{6/7}5^{3/7}N^{\gamma-1/7}},
\end{equation}
for the linear, the nonlinear and the quadratic spin squeezing respectively.
 Equations~(\ref{eq:bal_all}) show that for a linear measurement, a collective ballistic dephasing with $\gamma\leq 1/3$ does not change the best noiseless metrological gain. This is also true for a nonlinear measurement if $\gamma\leq 1/5$ and for a quadratic measurement if $\gamma \leq 1/7$.

\subsubsection{MAI measurements}
We have shown { in section \ref{sec:MAI}} that the MAI method allows, with an appropriate value of $\alpha$, to reproduce all the scaling laws for the linear, nonlinear and the quadratic spin squeezing in the noiseless case. To show that this observation can be extended to realistic scenarios, we identify the limitations of the MAI metrological gain~(\ref{eq:MAIopt}) in the presence of ballistic dephasing\footnote{{The metrological gain is given by Eq.~(\ref{eq:xireex}) with $\vec{X}=\hat{U}_{\tau}^{\dagger}\hat{\vec{S}}\hat{U}_{\tau}$. The elements of the commutator and the covariance matrices including the average over the random variable $D$
\begin{align}
C_{kl}&=-i\int dD p(D)\langle\psi_t|[\hat{S}_k,e^{i\hat{H}_{\rm bal}\tau/\hbar}\hat{S}_le^{-i\hat{H}_{\rm bal}\tau/\hbar}]|\psi_t\rangle 
\label{eq:C}\\
\Gamma_{kl}&=\frac{1}{2}\int dD p(D)\langle\psi_{t}|e^{i\hat{H}_{\rm bal}\tau/\hbar}\{\hat{S}_k,\hat{S}_l\}e^{-i\hat{H}_{\rm bal}\tau/\hbar}|\psi_{t}\rangle
-\Pi_{j=l,k}\int dD p(D)\langle\psi_t|e^{i\hat{H}_{\rm bal}\tau/\hbar}\hat{S}_j e^{-i\hat{H}_{\rm bal}\tau/\hbar}|\psi_t\rangle
 \label{eq:gam}
\end{align}
where we take $\chi \tau=-\chi t$ as before and $D \tau=D t$, are given in~\ref{C}. 
}}.
\begin{figure}[tb]
    \centering
    \includegraphics[width=0.9\textwidth]{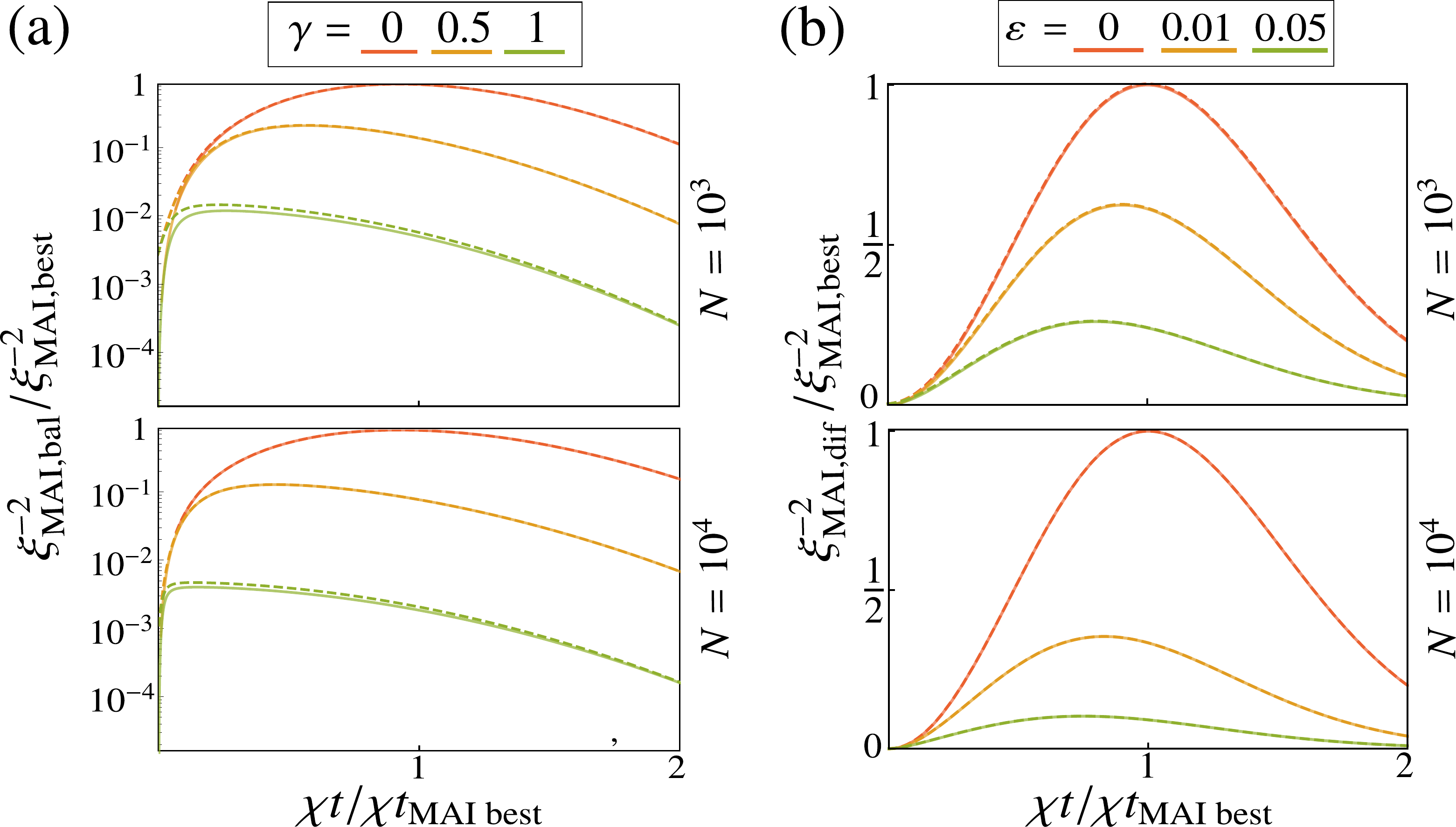}
    \caption{{ Metrological gain as a function of time using the measurement-after-interaction strategy in presence of decoherence. The atom number is $N=10^3$ (top row) and $10^4$ (bottom row). (a)  Ballistic dephasing with $\gamma=0,0.5,1$ and $\epsilon=0.05$. Solid lines are the analytic formulas (\ref{eq:MAIxibal}) for $\xi^{-2}_{\rm MAI,bal}$ in the limit of large $N$, and dashed lines are exact results. (b) Diffusive dephasing with $\varepsilon=0,0.01,0.05$. Solid lines are the analytical predictions  (\ref{eq:MAIxidif}) for $\xi^{-2}_{\rm MAI,dif}$ in the large $N$ limit, and dashed lines the exact result. }
}
    \label{ballistic_diffusive}
\end{figure}
The dominant effect of this random dephasing process is to increase, in a ballistic way, i.e., quadratically in $\chi t$, the variance of the optimal measurement observable $\approx\hat{S}_y$. { Indeed, for a small $\epsilon$, large $N$, and $\chi t\leq 1/\sqrt{N}$, after the second one-axis twisting evolution in presence of ballistic noise we obtain}
\begin{align}\label{eq:balnoise}
(\Delta\hat{S}_y)^2_{\mathrm{bal}}=\frac{N}{4}\left[1+4\epsilon N^{1+\gamma}(\chi t)^2+\mathcal{O}(\chi t)^4\right].
\end{align}
This decreases the metrological gain (\ref{eq:MAIxi}) by a factor $(1+4\epsilon N^{1+\gamma}(\chi t)^2)^{-1}$
\begin{align}\label{eq:MAIxibal}
\left(\xi_{\rm MAI,bal}^{-2}(\chi t)\right)_{N\to\infty}=\frac{N^2(\chi t)^2e^{-N(\chi t)^2}}{1+4\epsilon N^{1+\gamma}(\chi t)^2}.
\end{align}
This expression is compared to the exact result in Fig.~\ref{ballistic_diffusive}(a) for different values of $\gamma$ and $N$.

{ Using Eq.~(\ref{eq:MAIxibal}), we can deduce the scaling laws for large $N$ of the gain on time scales} $\chi t=\sigma N^{-\alpha}$ with $1\geq\alpha \geq 1/2$:
\begin{align}\label{eq:MAIopt0}
\xi^{-2}_{\mathrm{MAI,bal}}=
\begin{cases}
\frac{\sigma^2 N^{2-2\alpha}}{1+4\epsilon \sigma^2 N^{1+\gamma-2\alpha}}, &\: 1\geq\alpha > 1/2\\
&\\
\frac{\sigma^2 e^{-\sigma^2}N}{1+4\epsilon\sigma^2 N^{\gamma}}, &\:  \alpha = 1/2
\end{cases}.
\end{align}
We thus observe the existence of a critical value of $\alpha$
\begin{align}\label{eq:alpha_c}
\alpha_c=\frac{1+\gamma}{2},
\end{align}
such that for $\alpha\geq\alpha_c$ { the gain (\ref{eq:MAIopt0}) corresponds to the noiseless scaling law (\ref{eq:MAIopt}), while for $\alpha<\alpha_c$}, the gain is affected by the dephasing and becomes independent of $\alpha$. 
\begin{align}\label{eq:MAIblong}
\xi^{-2}_{\mathrm{MAI,bal}}\simeq
 \frac{1}{4\epsilon } N^{1-\gamma} \,.
\end{align}
A maximization of $\xi^{-2}_{\rm MAI,bal}$ over $\sigma$ and $\alpha$ allows us to find, for a given $\gamma$, the scaling law of the best metrological gain and the corresponding time.  For $\gamma=0$, we obtain
\begin{equation}
\chi t_{\rm MAI,bal,best}\simeq(1-2\epsilon)N^{-1/2} \qquad ; \qquad
\xi^{-2}_{\rm MAI,bal,best}\simeq\frac{\epsilon}{e}N,
\end{equation} 
while for $\gamma\neq 0$, the scaling law (\ref{eq:MAIblong}) represents the maximum  metrological gain. This is achieved exactly at the critical point $\alpha_{c}$, as well as by all longer times. 
By including first finite size corrections to (\ref{eq:MAIopt0}), we obtain
\begin{align}\label{eq:xibalcor}
\xi^{-2}_{\rm MAI,bal}\simeq\frac{\sigma^2 N^{2-2\alpha}}{1+4\epsilon \sigma^2 N^{1+\gamma-2\alpha}}[1-\sigma^2 N^{1-2\alpha}+\frac{\sigma^4}{4} N^{2-4\alpha}].
\end{align}
A maximization over $\alpha$ and $\sigma$ of (\ref{eq:xibalcor}), shows that $\xi^{-2}_{\rm MAI,bal}$ attains its maximal value (\ref{eq:MAIblong}) at $\chi t=(4\epsilon)^{-1/4}N^{-1/2-\gamma/4}$. 
In general, for a desired value of $\alpha$, Eq.~(\ref{eq:alpha_c}) sets a maximal tolerable level of ballistic dephasing noise $\gamma=2\alpha-1$ up to which the noiseless metrological gain is not affected by the ballistic dephasing. As we { already} observed in Eqs.~(\ref{eq:bal_all}), for the linear spin squeezing where the best time corresponds to $\alpha=2/3$, the tolerable noise level is $\gamma=1/3$; for the nonlinear squeezing where $\alpha=3/5$, this is given by $\gamma=1/5$ and it is given by $\gamma=1/7$ for the quadratic spin squeezing where $\alpha=4/7$. We thus demonstrate, as in the noiseless case, that the MAI technique allows to reproduces all the scaling laws of the metrological gain of different squeezing strategies also in the presence of ballistic dephasing.  





\subsection{Diffusive dephasing}
The OAT evolution in some experimental realizations is accompanied by collective spin fluctuations of diffusive nature. To describe these fluctuations, we consider a collective dephasing process at a rate $\gamma_C$ where the dynamics is governed by the master equation~\cite{Tannoudji,Breuer} with the Lindblad operator $\hat{L}=\hat{S}_z$
\begin{equation}
\frac{\partial \hat{\rho}}{\partial t}=-\frac{i}{\hbar}[\hat{H},\hat{\rho}]+\gamma_C\mathcal{L}[\hat{\rho}],
\end{equation}
where, $\mathcal{L}[\hat{\rho}]=\hat{S}_z\hat{\rho}\hat{S}_z-\frac{1}{2}\{\hat{S}_z^2,\hat{\rho}\}$ and $\hat{H}$ is the noiseless OAT Hamiltonian $\hat{H}=\hbar\chi\hat{S}_z^2$. Starting from the coherent spin state $\hat{\rho}_0=|\psi_0\rangle\langle\psi_0|$ where $|\psi_0\rangle$ is given by (\ref{eq:CCS}), the evolution of the system is given by 
\begin{equation}
    \hat{\rho}(t)=e^{\varepsilon\chi t\mathcal{L}}[\hat{U}_t \hat{\rho}_0\hat{U}_t^{\dagger}] \quad \textrm{with} \quad \varepsilon=\frac{\gamma_C}{\chi},
\end{equation}
where we used the fact that $[\hat{H},\hat{L}]=0$. Using $\mathcal{L}^{\dagger}=\mathcal{L}$, the expectation value of any operator $\hat{A}$ can be obtained from the adjoint master equation~\cite{Breuer} as
\begin{equation}\label{eq:Exp}
     \langle\hat{A}\rangle=\mathrm{tr}\{\hat{A}\hat{\rho}(t)\}=\mathrm{tr}\{e^{\varepsilon\chi t\mathcal{L}}[\hat{A}]\hat{U}_t \hat{\rho}_0\hat{U}_t^{\dagger}\}.
\end{equation}
These expressions can then be inferred from the noiseless expectation values by explicitly determining the transformed operator $e^{\varepsilon\chi \tau\mathcal{L}}[\hat{A}]$.
\subsubsection{Linear, nonlinear, and quadratic spin observables}

In the limit $N\gg 1$, the metrological gain of the linear, the nonlinear and the quadratic spin squeezing in the presence of a diffusive dephasing for $\chi t<1/\sqrt{N}$ are obtained using the same steps as before\footnote{{ The elements of the commutator and covariance matrices to be used in the metrological gain~(\ref{eq:xireex}) now read
\begin{align}
C_{kl}&=-i\langle\psi_t|e^{\varepsilon\chi t\mathcal{L}_C}\left[[\hat{S}_k,\hat{X}_l]\right]|\psi_t\rangle \qquad ; \qquad
\Gamma_{kl}=\frac{1}{2}\langle\psi_t|e^{\varepsilon\chi t\mathcal{L}_C}[\{\hat{X}_k,\hat{X}_l\}]|\psi_t\rangle
-\langle\psi_t|e^{\varepsilon\chi t\mathcal{L}_C}[\hat{X}_k]|\psi_t\rangle 
\langle\psi_t|e^{\varepsilon\chi t\mathcal{L}_C}[\hat{X}_l]|\psi_t\rangle,
\end{align}
where $|\psi_t\rangle=e^{-i\chi t\hat{S}_z^2}|\psi_0\rangle$. Their analytical expression is given in~\ref{D}.}} and read
\begin{equation}\label{eq:dif}
\xi^{-2}_{\rm dif}(\chi t)\simeq\frac{N^2(\chi t)^2}{1+M+\varepsilon N\chi t},
\end{equation}
with the appropriate expression of $M$, which is given in Sec.~\ref{sec:QFI}. To maximize over $\chi t$ in the limit of large $N$ at fixed $\varepsilon$, where 
 $\varepsilon\gg1/(N \chi t_{best})$, we can approximate (\ref{eq:dif}) as
\begin{equation}\label{eq:difapp}
\xi^{-2}_{\rm dif}(\chi t)\simeq\frac{N^2(\chi t)^2}{M+\varepsilon N\chi t}.
\end{equation}
We then find { the best time} for the linear, nonlinear and the quadratic squeezing in presence of the diffusive dephasing 
\begin{equation}\label{dift_all}
\chi t_{\mathrm{best,L (dif)}}\simeq\left(\frac{3\varepsilon}{2}\right)^{1/5}N^{-3/5}\quad ; \quad
\chi t_{\mathrm{best,NL (dif)}}\simeq\left(\frac{5\varepsilon}{4}\right)^{1/9}3^{1/3}N^{-5/9}\quad ; \quad
\chi t_{\mathrm{best,Q (dif)}}\simeq\left(\frac{7}{3}\right)^{1/13}\frac{5^{3/13}\varepsilon^{1/13}}{2^{2/13}}N^{-7/13}\,,
\end{equation}
and corresponding best metrological gain 
\begin{equation}\label{difxi_all}
\xi_{\mathrm{best,L (dif)}}^{-2}\simeq\frac{2\times 3^{1/5}}{5}\left(\frac{2}{\varepsilon}\right)^{4/5} N^{2/5}\quad ; \quad
\xi_{\mathrm{best,NL (dif)}}^{-2}\simeq\frac{4}{3}\frac{2^{7/9}5^{1/9}}{3^{5/3}\varepsilon^{8/9}}N^{4/9} \quad ; \quad
\xi_{\mathrm{best,Q (dif)}}^{-2}\simeq \frac{2}{13}\frac{2^{11/13}3^{12/13}5^{3/13}7^{1/13}}{\varepsilon^{12/13}} N^{6/13}.
\end{equation}
For a linear measurement, the Eqs.~(\ref{dift_all}) and~(\ref{difxi_all}) confirm the optimal scaling laws $\chi t_{\rm best}\propto N^{-3/5}$ and $\xi^{-2}_{\rm best}\propto N^{2/5}$ found in the presence of diffusive dephasing due to cavity losses in cavity induced spin squeezing~\cite{PawlowskiEPL2016,MonikaPRA2010,LerouxPRA2012}.

\subsubsection{MAI measurements}
For the MAI measurement, the quantum gain is again given by (\ref{eq:xireex}) with $\vec{X}=\hat{U}_{\tau}^{\dagger}\hat{\vec{S}}\hat{U}_{\tau}$ with the following elements of $C$ and $\Gamma$ 
\begin{align}
C_{kl}&=-i\langle\psi_t|e^{\varepsilon\chi t\mathcal{L}_C}\left[[\hat{S}_k,\hat{U}_{\tau}^{\dagger}e^{\varepsilon\chi\tau\mathcal{L}_C}[\hat{S}_l]\hat{U}_{\tau}]\right]|\psi_t\rangle,\label{eq:MAIdifc} \\
 \Gamma_{kl}&=\frac{1}{2}\langle\psi_t|e^{\varepsilon\chi t\mathcal{L}_C}\left[\hat{U}_{\tau}^{\dagger}e^{\varepsilon\chi\tau\mathcal{L}_C}[\{\hat{S}_k,\hat{S}_l\}]\hat{U}_{\tau}]\right]|\psi_t\rangle
- \Pi_{j=k,l}
\langle\psi_t|e^{\varepsilon\chi t\mathcal{L}_C}\left[\hat{U}_{\tau}^{\dagger}e^{\varepsilon\chi\tau\mathcal{L}_C}[\hat{S}_j]\hat{U}_{\tau}]\right]|\psi_t\rangle
\label{eq:MAIdifgam}
\end{align}
The analytical expressions of (\ref{eq:MAIdifc}) and (\ref{eq:MAIdifgam}) are given in Appendix~\ref{D}. Taking the optimization~(\ref{eq:tauopt}) into account, we replace $\tau=-t$. The variance of the optimal measurement observable $\hat{S}_y$ here increases as
\begin{align}\label{eq:balnoise}
(\Delta\hat{S}_y)^2_{\mathrm{dif}}=\frac{N}{4}\left[1+2 \varepsilon N \chi t+\mathcal{O}(\chi t)^2\right],
\end{align}
showing a diffusive behavior, i.e., linear in $\chi t$. This limits the quantum metrological gain of the MAI technique (\ref{eq:MAIxi}) and indeed we find for $\chi t\leq 1/\sqrt{N}$
\begin{equation}\label{eq:MAIxidif}
\xi^{-2}_{\rm MAI,dif}(\chi t)=\frac{N^2(\chi t)^2 e^{-N(\chi t)^2}}{1+2 \varepsilon N \chi t}.
\end{equation}
This expression is represented and compared to exact results in Fig.~\ref{ballistic_diffusive}(b) for varying $\varepsilon$ and $N$.

Again, we obtain the scaling laws of the metrological gain on the time scales $\chi t=\sigma N^{-\alpha}$ in the limit of large $N$
\begin{align}\label{xidif}
\left(\xi^{-2}_{\mathrm{MAI,dif}}\right)_{N\to\infty}=
\begin{cases}
\frac{\sigma}{2\varepsilon}N^{1-\alpha}, &\quad 1\geq\alpha > 1/2\\
&\\
\frac{\sigma e^{-\sigma^2}}{2\varepsilon}N^{1/2}, &\quad \alpha = 1/2
\end{cases}.
\end{align}
Due to the diffusive dephasing, the scaling law of the metrological gain for the MAI method passes from $\xi^{-2}_{\rm MAI}\propto N^{2-2\alpha}$ to $\xi^{-2}_{\rm MAI}\propto N^{1-\alpha}$ for a given $\alpha$. As expected, the scaling for the MAI method reproduces the scaling laws (\ref{difxi_all}) for the states prepared at the times (\ref{dift_all}).

For $1/2\leq\alpha \leq 1$, an optimization of (\ref{xidif}) over $\alpha$ and $\sigma$ gives us the best metrological gain and the corresponding time for $N\gg 1$
\begin{equation}
\chi t_{\rm MAI,dif,best}=\frac{1}{\sqrt{2}}N^{-1/2} \qquad ; \qquad
\xi^{-2}_{\rm MAI,dif,best}=\frac{N^{1/2}}{\sqrt{8\varepsilon^2 e}}.
\end{equation}
This analytically confirms a result that was obtained numerically in Ref.~\cite{HammererQuantum2020}.

\subsection{Unified expression}
Taking $e^{-N(\chi t)^2}\approx 1$ for $\chi t<1/\sqrt{N}$ and $N\gg 1$, Eqs.~(\ref{eq:bal}), (\ref{eq:MAIxibal}), (\ref{eq:dif}) and~(\ref{eq:MAIxidif}) show that in the presence of decoherence, the metrological gain can again be written { with an unified expression}:
\begin{align}\label{eq:unif}
\xi^{-2}(\chi t)\simeq\frac{F_Q/N}{1+M+B},
\end{align}
where $B_{\rm bal}=\epsilon N^{1+\gamma}(\chi t)^2$  and $B_{\rm dif}=\varepsilon N\chi t$ describes the loss of sensitivity due to ballistic and diffusive dephasing, for the linear, nonlinear and quadratic measurements. In the case of an MAI measurement, the nonlinear OAT evolution is effectively twice as long, which increases the effect of the decoherence. This effect can be easily accounted for by replacing $\chi t$ by $2\chi t$ in the case of MAI for the decoherence terms, leading to $B_{\rm bal}=4\epsilon N^{1+\gamma}(\chi t)^2$ and $B_{\rm dif}=2\varepsilon N\chi t$. The result~(\ref{eq:unif}) allows us to obtain, in a simple way, the scaling laws and optimal times in all cases discussed above.

\section{Particle losses}
Up to now we have considered dephasing processes perturbing the coherent evolution with the OAT Hamiltonian. In this { last} section we will explore the limitiations imposed by particle losses to the linear, nonlinear and quadratic spin squeezing.
\subsection{ Loss model}
For convenience we write here the collective spin components using { the creation $\hat{c}_a^{\dagger}$ ($\hat{c}_b^{\dagger}$) and the annihilation $\hat{c}_a$ ($\hat{c}_b$)} operators corresponding to the mode $a$ ($b$) respectively~:
\begin{align}
\hat{S}_x =\frac{\hat{c}_a^{\dagger}\hat{c}_b+\hat{c}_b^{\dagger}\hat{c}_a}{2}\; , \; \hat{S}_y =\frac{\hat{c}_a^{\dagger}\hat{c}_b-\hat{c}_b^{\dagger}\hat{c}_a}{2i}\; , \; \hat{S}_z =\frac{\hat{c}_a^{\dagger}\hat{c}_a-\hat{c}_b^{\dagger}\hat{c}_b}{2}\,,
\end{align}
{ and we} introduce the phase state 
\begin{align}\label{eq:phstate}
|\varphi\rangle_N\equiv\frac{1}{\sqrt{N !}}\left(\frac{e^{i \varphi }\hat{c}_a^{\dagger}+e^{-i \varphi }\hat{c}_b^{\dagger}}{\sqrt{2}}\right)^N|0\rangle\,.
\end{align}
Note that $|\varphi=0\rangle_N$ corresponds to the coherent spin state (\ref{eq:CCS}) with $\langle\varphi=0| \hat{N}_l|\varphi=0\rangle|_{l=a,b}=N/2$ where $\hat{N}_l=\hat{c}_l^{\dagger}\hat{c}_l$ is the operator of number of particles in the mode $l$. The presence of $m$-body losses, in addition to the one-axis-twisting dynamics $\hat{H}=\hbar\chi\hat{S}_z^2$, can be described by the master equation \cite{SinatraEuro1998},
\begin{align}
\frac{\partial\hat{\rho}}{\partial t}=-\frac{i}{\hbar}[\hat{H},\hat{\rho}]+\sum_{l=a,b}\gamma_l^{(m)}\left([\hat{c}_l]^m \hat{\rho} [\hat{c}_l^{\dagger}]^m-\frac{1}{2}\left\{[\hat{c}_l]^m [\hat{c}_l^{\dagger}]^m,\hat{\rho}\right\}\right)
\end{align}
where $\gamma^{(m)}_l$ is the $m$-body loss rate in the mode $l$. This evolution can be equivalently represented in terms of the Monte-Carlo wave function formalism~\cite{MolmerOpt1993}. In this point of view, the system is described by a wave function whose evolution is generated by an effective Hamiltonian $\hat{H}_{\mathrm{eff}}$ in time intervals of duration $\tau_j$ separated by random quantum jumps, described by the jump operators $\hat{J}_l^{(m)}$, at times $t_j$:
\begin{align}\label{eq:Heff}
\hat{H}_{\rm eff}=\hat{H}-\frac{i\hbar}{2}\sum_{l=a,b}\hat{J}_l^{(m) \dagger}\hat{J}_l^{(m)}\quad \textrm{with}\quad \hat{J}_l^{(m)}=\sqrt{\gamma_l^{(m)}} [\hat{c}_l]^m.
\end{align}
As long as the fraction of lost particles is weak we can approximate the effective Hamiltonian (\ref{eq:Heff}) by \cite{SinatraEuro1998}
\begin{align}\label{eq:approx}
\hat{H}_{\rm eff}=\hat{H}-\frac{i\hbar}{2}\lambda,
\end{align}
where $\lambda=\sum_{l=a,b}\lambda_l$ with $\lambda_l=\gamma_l^{(m)}\langle \hat{c}_l^{\dagger m}\hat{c}_l^{m}\rangle_{\psi_0}$. For simplicity, we restrict, in the following, to the symmetric case where $\gamma_a^{(m)}=\gamma_b^{(m)}=\gamma^{(m)}$. We assume that the system is initially in the phase state (\ref{eq:phstate}) with $\varphi=0$. In a particular Monte-Carlo realization with $k$ quantum jumps, each resulting in $m$-body losses in the mode $l_i=a,b$ at times $t_i$ with $i=1,...,k$, the state of the system at time $t$ is given by
\begin{align}\label{eq:StateLoss}
|\psi(t)\rangle =\mathcal{N} e^{-\frac{i}{\hbar}\hat{H}_{\rm eff}(t-t_k)}\hat{J}_{l_k}e^{-\frac{i}{\hbar}\hat{H}_{\rm eff}(t_k-t_{k-1})}...\hat{J}_{l_1}e^{-\frac{i}{\hbar}\hat{H}_{\rm eff}t_1}|\varphi=0\rangle_N
\end{align}
with $\mathcal{N}$ a normalization constant. By using the identity
\begin{align}
\hat{c}_l^m f(\hat{N}_a,\hat{N}_b)=f(\hat{N}_a+m\delta_{l,a},\hat{N}_b+m\delta_{l,b})\hat{c}_l^m
\end{align}
 for $l=a,b$ and the properties of phase states (\ref{eq:phstate})
\begin{equation}
\hat{c}_l|\varphi\rangle_N=\sqrt{\frac{N}{2}}e^{i\varphi(\delta_{l,a}-\delta_{l,b})}|\varphi\rangle_{N-1} \qquad ; \qquad
e^{-i\alpha(\hat{N}_a-\hat{N}_b)}|\varphi\rangle_N=|\varphi+\alpha\rangle_N,
\end{equation}
we can show that, in the approximation (\ref{eq:approx}), { the state (\ref{eq:StateLoss}) for a particular Monte-Carlo realization can be written as a shifted phase state with less particles, evolved with the one-axis-twisting hamiltonian. In terms of a normalization factor $F(t)$ and a random relative phase shift $D$~:} 
\begin{equation}
|\psi(t)\rangle= F(t) e^{-i\chi t \hat{S}_z^2} |D\rangle_{N-mk} \qquad ; \qquad
 D=m\sum_{i=1}^k \chi t_i\left( \delta_{b,c_i}-\frac{1}{2}\right).
\end{equation}
The expectation value of any operator $\hat{O}$ can be calculated by averaging the single realization mean value
\begin{equation}\label{eq:SinglReal}
\langle\psi(t)|\hat{O}|\psi(t)\rangle = \vphantom{}_{N-mk} \langle D|e^{i\chi t \hat{S}_z^2}\hat{O}e^{-i\chi t \hat{S}_z^2} |D\rangle_{N-mk}
\end{equation}
over all Monte-Carlo realizations, that is to average (\ref{eq:SinglReal}) over the random variables $k$, $t_i$ and $\delta_{b,c_i}$~\cite{SinatraEuro1998}. This allows us to analytically calculate the commutator and the covariance matrices (given in \ref{E}) and thus to obtain the quantum metrological gain (\ref{eq:Xilambda}) corresponding to the squeezing of a linear, nonlinear and quadratic spin observable in presence of $m$-body losses. In Fig. \ref{Xidif} (b), we compare the analytical metrological gain for the linear, nonlinear and quadratic spin squeezing in presence of one-body losses { in the approximation (\ref{eq:approx}), valid for the loss of a small fraction of the particles, to the exact numerical Monte-Carlo simulation with the effective Hamiltonian (\ref{eq:Heff}).}
\subsection{Scaling laws of the linear, nonlinear and quadratic spin squeezing}
\begin{figure}[htb]
    \includegraphics[width=0.9\textwidth]{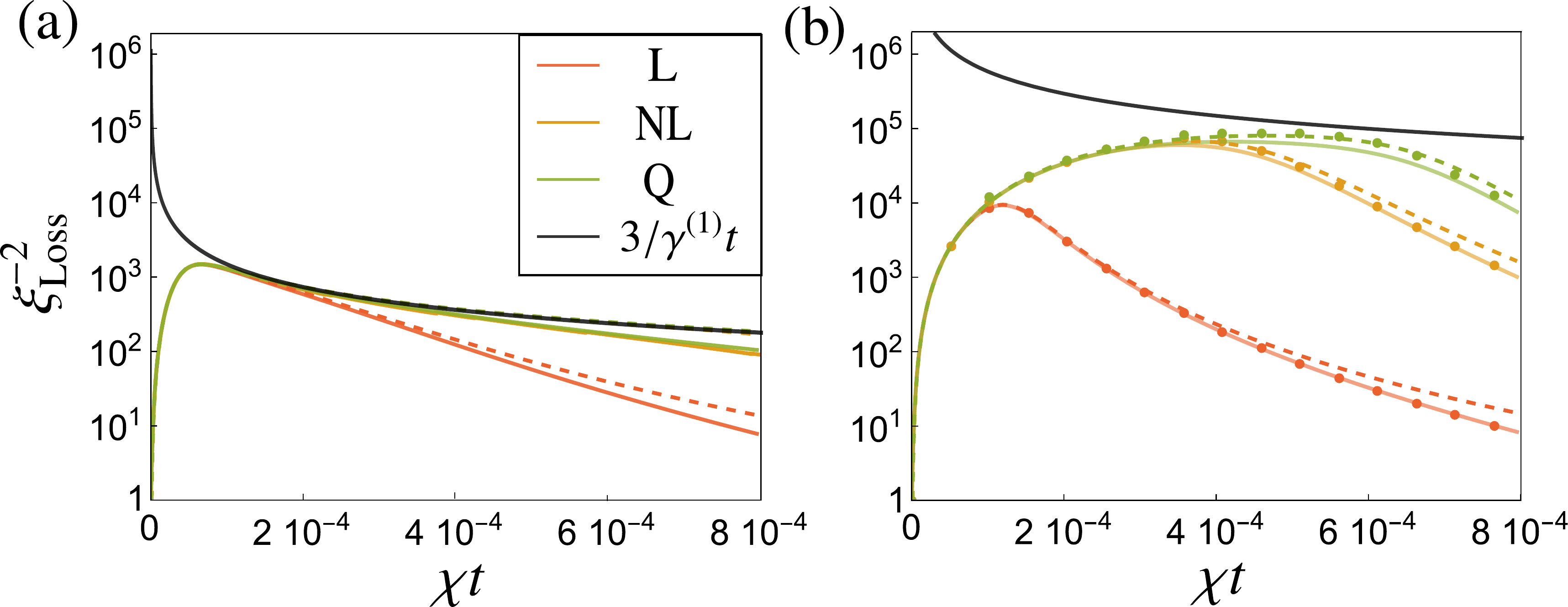}
    \caption{Linear, nonlinear and quadratic metrological gain for $N=10^6$ as a function of time in presence of one-body losses with (a) {$\gamma^{(1)}/\chi=20$ and (b)  $\gamma^{(1)}/\chi=0.05$} compared to the limit (\ref{eq:lim}). The metrological gain at the limit $N\gg 1$ given by (\ref{eq:LosL}), (\ref{eq:LosNL}) and (\ref{eq:LosQ}) are represented in dashed lines. Points in (b) are results of numerical Monte-Carlo simulation with 600 realizations.}
    \label{Xidif}
\end{figure}

Let us focus on the  case of 1-body losses ($m=1$) with a loss rate $\gamma^{(1)}$. To obtain the best metrological gain of the linear spin squeezing in the limit of large $N$ , { we use the best linear squeezing time to introduce an auxiliary dimensionless variable ${ r=N^{-1/3}}$ and rescale} the time as $\chi t=\theta r^{2}$. By expanding the linear metrological gain $\xi^{-2}_{\rm L}$ for $r\ll 1$ and $\gamma^{(1)}/\chi$ constant, we obtain
\begin{align}\label{eq:LosL}
\left(\xi^{-2}_{\rm L}(t)\right)_{N\to\infty}=\frac{N^2(\chi t)^2}{1+N^4(\chi t)^6/6+(\gamma^{(1)} t/3) N^{2}(\chi t)^2}.
\end{align}
Similarly, using the best nonlinear squeezing time (\ref{eq:bestinf}), we set ${ r=N^{-1/5}}$ and we rescale the time as $\chi t=\theta r^{3}$ to obtain
\begin{align}\label{eq:LosNL}
\left(\xi^{-2}_{\rm NL}(t)\right)_{N\to\infty}=\frac{N^2(\chi t)^2}{1+N^6(\chi t)^{10}/270+(\gamma^{(1)} t/3) N^{2}(\chi t)^2}.
\end{align}
For the quadratic squeezing, after setting ${ r=N^{-1/7}}$, rescaling the time as $\chi t=\theta r^{4}$ we obtain 
\begin{align}\label{eq:LosQ}
\left(\xi^{-2}_{\rm Q}(t)\right)_{N\to\infty}=\frac{N^2(\chi t)^2}{1+N^8(\chi t)^{14}/875+(\gamma^{(1)} t/3) N^{2}(\chi t)^2}.
\end{align}
By comparing the equations (\ref{eq:LosL})-(\ref{eq:LosQ}) to the equation (\ref{eq:bal}), we deduce that the effect of one-body losses is equivalent to the ballistic dephasing effect discussed in paragraph \ref{Bal} with $\gamma =1$ and $\epsilon = \gamma^{(1)}t/3$ where $\gamma^{(1)}t$ corresponds to the lost fraction of atoms at time $t$. { For the three measurement strategies, the metrological gain in the large $N$ limit, taken at constant lost fraction at $t_{\rm best}$, is then limited by the fraction of lost atoms}
\begin{align}\label{eq:lim}
\xi^{-2} = \frac{3}{\gamma^{(1)} t}.
\end{align} 
We then conclude, as shown in Fig. \ref{Xidif}. that for a fixed atom number $N$, a nonlinear measurement can enhance the linear metrological gain as long as $3/(\gamma^{(1)}t_{\rm L,best}) > \xi^{-2}_{\rm L, best}$. Such a regime can be reached as long as the $1$-body loss rate $\gamma^{(1)}$ is not too large (Fig. \ref{Xidif} b).
\section{Conclusion}
We have analytically found the scaling laws of the metrological gain in the limit of large atom numbers $N$ for the squeezing of nonlinear  spin observables. For the effective measurement of a nonlinear spin observable, we have identified the measurement-after-interaction technique that consists in adding a second nonlinear evolution before the direct measurement of a linear spin observable as a feasible possibility. This method indeed gives rise to a general scaling law for the metrological gain that continuously connects the different cases of measurement strategies based on linear and second-order spin observables. 

We have identified the limits imposed by two different models of decoherence, describing dominant decoherence processes in different physical realizations of the one-axis-twisting evolution. In the presence of ballistic collective dephasing, our results predict, in the thermodynamic limit, an abrupt change of the metrological gain at a critical preparation time that depends on the noise. This transition determines the longest state preparation time by one-axis-twisting for which the quantum scaling enhancement can be sustained in the presence of dephasing. Below this critical evolution time, the quantum gain is not affected by decoherence. In contrast, for diffusive dephasing, the scaling law corresponds to the square root of the gain in the noiseless case, independently of the preparation time. Finally, in the presence of particle losses, the best linear, nonlinear and quadratic spin squeezing are limited by the fraction of lost particles at the best squeezing time.

Our work analytically identifies the maximally achievable quantum sensitivity gain offered by the squeezing of nonlinear spin observables during a realistic one-axis-twisting evolution with an arbitrary number of atoms. As a function of the chosen measurement strategy, we identify optimal rotation directions, measurement observables and preparation times. These results may serve as a guide for designing feasible strategies for achieving high quantum enhancements in quantum phase estimation protocols with a relatively large number of atoms.

\section*{Acknowledgment}
M.G. acknowledges funding from the LabEx ENS-ICFP: ANR-10-LABX-0010 / ANR-10-IDEX-0001-02 PSL, from MCIN / AEI for the project PID2020-115761RJ-I00, and support of a fellowship from `la Caixa” Foundation (ID 100010434) and from the European Union’s Horizon 2020 research and innovation programme under the Marie Sk\l{}odowska-Curie grant agreement No 847648, fellowship code LCF/BQ/PI21/11830025.

\appendix
\section{Covariance and commutator matrices  for  quadratic measurements}\label{A}
Here we provide the { non zero elements of the $2\times 4$ commutator matrix $C$ (\ref{eq:Cdef}) obtained after restricting the interferometric rotation direction to the $yz$ plane $\hat{\vec{S}}=(\hat{S}_y,\hat{S}_z)^T$, and the elements of the} symmetric $4\times 4$ covariance matrix $\Gamma$ (\ref{eq:gammadef}) for the family of accessible observables $\hat{\vec{X}}_{\rm Q}=(\hat{S}_y,\hat{S}_z,\frac{1}{2}\{\hat{S}_x,\hat{S}_z\},\frac{1}{2}\{\hat{S}_x,\hat{S}_y\})^T$ that corresponds to a quadratic (Q) measurement. 
The results for the family $\hat{\vec{X}}_{\rm NL}$, which corresponds to a nonlinear (NL) measurement,  are obtained by focusing only on the $2\times 3$ sub-matrix of $C$ and the $3\times 3$ sub-matrix of $\Gamma$.
\begin{align*}
    C_{12}&= \frac{N}{2}\cos^{N-1}{(\chi t)} \quad ; \qquad  \:\:
    C_{13}=\frac{N(N-1)}{8}(\cos^{N-2}{(2\chi t)}+1) \quad ; \quad \:\:
    C_{14}=-\frac{N(N-1)}{4}\sin{(\chi t)}\cos^{N-2}{(\chi t)}\\
    C_{21}&=-C_{12} \quad ; \qquad \qquad \qquad
    C_{23}=\frac{N(N-1)}{4}\sin{(\chi t)}\cos^{N-2}{(\chi t)} \quad ; \quad
     C_{24}=-\frac{N(N-1)}{4}\cos^{N-2}{(\chi t)}
\end{align*}
\begin{align*}
     \Gamma_{11}&=\frac{N(N+1)}{8}-\frac{N(N-1)}{8}\cos^{N-2}{(2\chi t)} \quad ; \qquad \:
    \Gamma_{12}= \frac{N(N-1)}{4}\sin{(\chi t)}\cos^{N-2}{(\chi t)},\\
    \Gamma_{13}&=\frac{N(N-1)(N-2)}{16}\sin{(2\chi t)}\cos^{N-3}{(2\chi t)}\quad ; \quad
    \Gamma_{14}=\frac{N(N-1)}{8}\cos^{N-1}{(\chi t)}
    -\frac{N(N-1)(N-2)}{32}\left(\cos^{N-3}{(3\chi t)}-\cos^{N-3}{(\chi t)}\right),\\
    \Gamma_{22}&=\frac{N}{4}\quad ; \quad
    \Gamma_{23}=\frac{N(N-1)}{8}\cos^{N-1}{(\chi t)}
    -\frac{N(N-1)(N-2)}{8}\sin^2{(2\chi t)}\cos^{N-3}{(\chi t)},\\
    \Gamma_{24}&=\frac{N(N-1)(N-2)}{16}\sin{(2\chi t)}\cos^{N-3}{(2\chi t)},\\
    \Gamma_{33}&=\frac{N(N-1)(N-2)}{32}(\cos^{N-2}{(2\chi t)}+1)
    -\frac{N(N-1)}{32}(\cos^{N-2}{(2\chi t)}-3)
    -\frac{N(N-1)(N-2)(N-3)}{32}\sin^2{(2\chi t)}\cos^{N-4}{(2\chi t)},\\
    \Gamma_{34}&=\frac{N(N-1)(N-2)}{16}\sin{(\chi t)}\cos^{N-2}{(\chi t)}
    -\frac{N(N-1)}{16}\sin{(\chi t)}\cos^{N-2}{(\chi t)}
    +\frac{N(N-1)(N-2)(N-3)}{64}\left(\sin{(\chi t)}\cos^{N-4}{(\chi t)}\right.\\
    &\left.\quad+\sin{(3\chi t)}\cos^{N-4}{(3\chi t)}\right),\\
    \Gamma_{44}&=-\frac{N(N-1)(N-2)(N-3)}{128}\cos^{N-4}{(4\chi t)}
    -\frac{1}{64}N(N-1)+\frac{1}{128}(N-1)(N+3)N^2.
\end{align*}
\section{Covariance and commutator matrices  for the MAI technique}\label{B}

{ Here we provide the non zero $2\times 2$ commutator matrix $C$ and the $2\times 2$ symmetric covariance matrix $\Gamma$ for the measurement-after-interction technique corresponding to the family of observables $\hat{\vec{X}}_{\rm MAI}=(\hat{U}_{\tau}^{\dagger}\hat{S}_y\hat{U}_{\tau},\hat{U}_{\tau}^{\dagger}\hat{S}_z\hat{U}_{\tau})^T$.}
\begin{align*}
    C_{11}&=\frac{N(N-1)}{4}\sin (\chi \tau)\left[\cos^{N-2}(\chi (\tau +2t))+\cos^{N-2}(\chi \tau )\right]\quad ; \quad
    C_{12}=\frac{N}{2}\cos^{N-1}{(\chi t)}\quad ; \quad
    C_{21}=-\frac{N}{2}\cos^{N-1}{(\chi (t+\tau))},\\
     \Gamma_{11}&=\frac{N(N+1)}{8}-\frac{N(N-1)}{8}\cos^{N-2}(2\chi(t+\tau))\quad ; \quad
    \Gamma_{12}=\frac{N(N-1)}{4}\cos^{N-2}(2\chi(t+\tau))\sin(\chi(t+\tau))\quad ; \quad
    \Gamma_{22}=\frac{N}{4}.
\end{align*}
\section{Covariance and commutator matrices in the presence of ballistic dephasing}\label{C}
In the following, we provide { the expressions of the commutator and the covariance matrices considered in \ref{A} and \ref{B}, including a ballistic dephasing in the state preparation (and measurement for MAI):
\begin{align}
C_{kl}&=-i\int e^{-\frac{D^2}{2\langle D^2\rangle}}\langle\psi_t|[\hat{S}_k,\hat{X}_l]|\psi_t\rangle dD,\\
\nonumber\Gamma_{kl}&=\frac{1}{2}\int e^{-\frac{D^2}{2\langle D^2\rangle}}\langle\psi_t|\{\hat{X}_k,\hat{X}_l\}|\psi_0\rangle dD 
-\int e^{-\frac{D^2}{2\langle D^2\rangle}}\langle\psi_t|\hat{X}_k |\psi_t\rangle dD\int e^{-\frac{D^2}{2\langle D^2\rangle}}\langle\psi_t|\hat{X}_l|\psi_t\rangle dD.
\end{align}
}
\begin{align*}
    C_{12}&= \frac{N}{2}e^{-\frac{1}{2}(\chi t)^2\langle D^2 \rangle}\cos^{N-1}{(\chi t)}\quad ; \hspace{4cm}
    C_{13}=\frac{N(N-1)}{8}\left(e^{-2(\chi t)^2\langle D^2 \rangle}\cos^{N-2}{(2\chi t)}+1\right)\,,\\
    C_{14}&=-\frac{N(N-1)}{4}e^{-\frac{1}{2}(\chi t)^2\langle D^2 \rangle}\sin{(\chi t)}\cos^{N-2}{(\chi t)},\quad ; \hspace{1.5cm}
    C_{21}=-C_{12}\,,\\
    C_{23}&=\frac{N(N-1)}{4}e^{-\frac{1}{2}(\chi t)^2\langle D^2 \rangle}\sin{(\chi t)}\cos^{N-2}{(\chi t)}\quad ; \hspace{2cm}
        C_{24}=-\frac{N(N-1)}{4}e^{-\frac{1}{2}(\chi t)^2\langle D^2 \rangle}\cos^{N-2}{(\chi t)},
\end{align*}
\begin{align*}
     \Gamma_{11}&=\frac{N(N+1)}{8}-\frac{N(N-1)}{8}e^{-2(\chi t)^2\langle D^2 \rangle}\cos^{N-2}{(2\chi t)}\quad ; \quad
    \Gamma_{12}= \frac{N(N-1)}{4}e^{-\frac{1}{2}(\chi t)^2\langle D^2 \rangle}\sin{(\chi t)}\cos^{N-2}{(\chi t)},\\
    \Gamma_{13}&=\frac{N(N-1)(N-2)}{16}e^{-2(\chi t)^2\langle D^2 \rangle}\sin{(2\chi t)}\cos^{N-3}{(2\chi t)},\\
    \Gamma_{14}&=\frac{N(N-1)}{8}e^{-\frac{1}{2}(\chi t)^2\langle D^2 \rangle}\cos^{N-1}{(\chi t)}
    -\frac{N(N-1)(N-2)}{32}\left(e^{-\frac{9}{2}(\chi t)^2\langle D^2 \rangle}\cos^{N-3}{(3\chi t)}
    -e^{-\frac{1}{2}(\chi t)^2\langle D^2 \rangle}\cos^{N-3}{(\chi t)}\right)
\end{align*}
\begin{align*}
    \Gamma_{22}&=\frac{N}{4}\quad ; \quad
    \Gamma_{23}=\frac{N(N-1)}{8}e^{-\frac{1}{2}(\chi t)^2\langle D^2 \rangle}\cos^{N-1}{(\chi t)}
    -\frac{N(N-1)(N-2)}{8}e^{-\frac{1}{2}(\chi t)^2\langle D^2 \rangle}\sin^2{(2\chi t)}\cos^{N-3}{(\chi t)},\\
    \Gamma_{24}&=\frac{N(N-1)(N-2)}{16}e^{-2(\chi t)^2\langle D^2 \rangle}\sin{(2\chi t)}\cos^{N-3}{(2\chi t)},\\
    \Gamma_{33}&=\frac{N(N-1)(N-2)}{32}(e^{-2(\chi t)^2\langle D^2 \rangle}\cos^{N-2}{(2\chi t)}+1)
    -\frac{N(N-1)}{32}(e^{-2(\chi t)^2\langle D^2 \rangle}\cos^{N-2}{(2\chi t)}-3)\\
    &\quad-\frac{N(N-1)(N-2)(N-3)}{32}e^{-2(\chi t)^2\langle D^2 \rangle}\sin^2{(2\chi t)}\cos^{N-4}{(2\chi t)},\\
    \Gamma_{34}&=\frac{N(N-1)(N-2)}{16}e^{-\frac{1}{2}(\chi t)^2\langle D^2 \rangle}\sin{(\chi t)}\cos^{N-2}{(\chi t)}
    -\frac{N(N-1)}{16}e^{-\frac{1}{2}(\chi t)^2\langle D^2 \rangle}\sin{(\chi t)}\cos^{N-2}{(\chi t)}\\
    &\quad+\frac{N(N-1)(N-2)(N-3)}{64}\left(e^{-\frac{1}{2}(\chi t)^2\langle D^2 \rangle}\sin{(\chi t)}\cos^{N-4}{(\chi t)}
    +\sin{(3\chi t)}e^{-\frac{9}{2}(\chi t)^2\langle D^2 \rangle}\cos^{N-4}{(3\chi t)}\right),\\
    \Gamma_{44}&=-\frac{N(N-1)(N-2)(N-3)}{128}e^{-8(\chi t)^2\langle D^2 \rangle}\cos^{N-4}{(4\chi t)}
    -\frac{1}{64}N(N-1)+\frac{1}{128}(N-1)(N+3)N^2.
\end{align*}
{ In the case of the MAI method we obtain~:}
\begin{align*}
    C_{11}&=\frac{N(N-1)}{4} e^{-\frac{1}{2}\chi^2(2t+\tau)^2\langle D^2 \rangle}\sin (\chi \tau)\cos^{N-2}(\chi (\tau +2t))
    +\frac{N(N-1)}{4} e^{-\frac{1}{2}(\chi\tau)^2\langle D^2 \rangle}\sin (\chi \tau)\cos^{N-2}(\chi \tau),\\
    C_{12}&=\frac{N}{2}e^{-\frac{1}{2}(\chi t)^2\langle D^2 \rangle}\cos^{N-1}{(\chi t)}\qquad , \qquad
    C_{21}=\frac{N}{2}e^{-\frac{1}{2}\chi^2 (t+\tau)^2\langle D^2 \rangle}\cos^{N-1}{(\chi (t+\tau))} \,. \\
     \Gamma_{11}&=\frac{N(N+1)}{8}-\frac{N(N-1)}{8}e^{-2\chi^2 (t+\tau)^2\langle D^2 \rangle}\cos^{N-2}(2\chi(t+\tau)) \:,\\
    \Gamma_{12}&=\frac{N(N-1)}{4}e^{-\frac{1}{2}\chi^2 (t+\tau)^2\langle D^2 \rangle}\cos^{N-2}(2\chi(t+\tau))\sin(\chi(t+\tau))
    \qquad , \qquad
    \Gamma_{22}=\frac{N}{4}.
\end{align*}

\section{Covariance and commutator matrices in presence of diffusive dephasing}\label{D}
{ Here we give the expressions of the commutator and the covariance matrices considered in \ref{A} and \ref{B}, including a collective diffusive dephasing in the state preparation (and measurement for MAI)}:
\begin{align*}
    C_{12}&= \frac{N}{2}e^{-\frac{\varepsilon}{2}\chi t}\cos^{N-1}{(\chi t)}\qquad , \hspace{3.2cm}
    C_{13}=\frac{N(N-1)}{8}\left(e^{-2\varepsilon\chi t}\cos^{N-2}{(2\chi t)}+1\right)\\
    C_{14}&=-\frac{N(N-1)}{4}e^{-\frac{\varepsilon}{2}\chi t}\sin{(\chi t)}\cos^{N-2}{(\chi t)} \qquad ; \qquad
    C_{21}=-C_{12},\\
    C_{23}&=\frac{N(N-1)}{4}e^{-\frac{\varepsilon}{2}\chi t}\sin{(\chi t)}\cos^{N-2}{(\chi t)} \qquad ; \qquad \quad
    C_{24}=-\frac{N(N-1)}{4}e^{-\frac{\varepsilon}{2}\chi t}\cos^{N-2}{(\chi t)}\\
    \Gamma_{11}&=\frac{N(N+1)}{8}-\frac{N(N-1)}{8}e^{-2\varepsilon\chi t}\cos^{N-2}{(2\chi t)} \qquad ; \qquad
    \Gamma_{12}= \frac{N(N-1)}{4}e^{-\frac{\varepsilon}{2}\chi t}\sin{(\chi t)}\cos^{N-2}{(\chi t)},\\
    \Gamma_{13}&=\frac{N(N-1)(N-2)}{16}e^{-2\varepsilon\chi t}\sin{(2\chi t)}\cos^{N-3}{(2\chi t)},\\
    \Gamma_{14}&=\frac{N(N-1)}{8}e^{-\frac{\varepsilon}{2}\chi t}\cos^{N-1}{(\chi t)}
    -\frac{N(N-1)(N-2)}{32}\left(e^{-\frac{9\varepsilon}{2}}\cos^{N-3}{(3\chi t)}
    -e^{-\frac{\varepsilon}{2}\chi t}\cos^{N-3}{(\chi t)}\right),\\
    \Gamma_{22}&=\frac{N}{4}\qquad ; \qquad
    \Gamma_{23}=\frac{N(N-1)}{8}e^{-\frac{\varepsilon}{2}\chi t}\cos^{N-1}{(\chi t)}
    -\frac{N(N-1)(N-2)}{8}e^{-\frac{\varepsilon}{2}\chi t}\sin^2{(2\chi t)}\cos^{N-3}{(\chi t)},\\
    \Gamma_{24}&=\frac{N(N-1)(N-2)}{16}e^{-2\varepsilon\chi t}\sin{(2\chi t)}\cos^{N-3}{(2\chi t)},\\
    \Gamma_{33}&=\frac{N(N-1)(N-2)}{32}(e^{-2\varepsilon\chi t}\cos^{N-2}{(2\chi t)}+1)
    -\frac{N(N-1)}{32}(e^{-2\varepsilon\chi t}\cos^{N-2}{(2\chi t)}-3)\\
    &\quad-\frac{N(N-1)(N-2)(N-3)}{32}e^{-2\varepsilon\chi t}\sin^2{(2\chi t)}\cos^{N-4}{(2\chi t)},\\
    \Gamma_{34}&=\frac{N(N-1)(N-2)}{16}e^{-\frac{\varepsilon}{2}\chi t}\sin{(\chi t)}\cos^{N-2}{(\chi t)}
    -\frac{N(N-1)}{16}e^{-\frac{\varepsilon}{2}\chi t}\sin{(\chi t)}\cos^{N-2}{(\chi t)}\\
    &\quad+\frac{N(N-1)(N-2)(N-3)}{64}\left(e^{-\frac{\varepsilon}{2}\chi t}\sin{(\chi t)}\cos^{N-4}{(\chi t)}
    +\sin{(3\chi t)}e^{-\frac{9\varepsilon}{2}\chi t}\cos^{N-4}{(3\chi t)}\right),\\
    \Gamma_{44}&=-\frac{N(N-1)(N-2)(N-3)}{128}e^{-8\varepsilon\chi t}\cos^{N-4}{(4\chi t)}
    -\frac{1}{64}N(N-1)+\frac{1}{128}(N-1)(N+3)N^2.
\end{align*}
For the MAI method, we obtain
\begin{align*}
    C_{11}&=\frac{N(N-1)}{4} e^{-\varepsilon[2|\chi t|+\frac{1}{2}|\chi\tau|]} \sin (\chi \tau)\cos^{N-2}(\chi (\tau +2t))
    +\frac{N(N-1)}{4} e^{- \frac{\varepsilon}{2}|\chi\tau|}\sin (\chi \tau)\cos^{N-2}(\chi \tau),\\
    C_{12}&=\frac{N}{2}e^{- \frac{\varepsilon}{2}|\chi t|}\cos^{N-1}{(\chi t)}\qquad , \qquad
    C_{21}=-\frac{N}{2}e^{-\frac{\varepsilon}{2}(|\chi t|+|\chi\tau|)}\cos^{N-1}{(\chi (t+\tau))},\\
     \Gamma_{11}&=\frac{N(N+1)}{8}-\frac{N(N-1)}{8}e^{-2\varepsilon(|\chi t|+|\chi\tau|)}\cos^{N-2}(2\chi(t+\tau)),\\
    \Gamma_{12}&=\frac{N(N-1)}{4}e^{-\frac{\varepsilon}{2}(|\chi t|+|\chi\tau|)}\cos^{N-2}(\chi(t+\tau))\sin(\chi(t+\tau)) \qquad , \qquad
    \Gamma_{22}=\frac{N}{4}.
\end{align*}

\section{Covariance and commutator matrices in presence of $m$-body losses}\label{E}
Here we provide the commutator matrix $C$ and the symmetric covariance matrix $\Gamma$ for the quadratic squeezing in the presence of $m$-body losses with loss rate $\gamma^{(m)}$ and the mean total number of $m$-body losses events per unit of time $\lambda$. As explained above, the commutator and the covariance matrices for the linear and nonlinear squeezing are obtained by restricting $C$ and $\Gamma$ to the first two lines and columns and to the first three lines and columns respectively:

\begin{align*}
    C_{12}&= \frac{N-m\lambda t\frac{\text{sinc}(m \chi t)}{\cos ^m(\chi t)}}{2}\cos^{N-1}{(\chi t)} e^{-\lambda  t \left(1-\frac{\text{sinc}(m \chi t)}{\cos ^m(\chi t)}\right)}\qquad , \qquad
    C_{13}=\frac{1}{8} F(0)+\frac{1}{8}F(2\chi t) \cos^{N-2}{(2\chi t)} e^{-\lambda  t \left(1-\frac{\text{sinc}(2m \chi t)}{\cos ^m(2\chi t)}\right)}:,\\
    C_{14}&=-\frac{1}{4}F(\chi t)\sin{(\chi t)}\cos^{N-2}{(\chi t)} e^{-\lambda  t \left(1-\frac{\text{sinc}(m \chi t)}{\cos ^m(\chi t)}\right)}
    \qquad , \qquad
    C_{21}=-C_{12}\:,\\
    C_{23}&=\frac{1}{4}F(\chi t)\sin{(\chi t)}\cos^{N-2}{(\chi t)} e^{-\lambda  t \left(1-\frac{\text{sinc}(m \chi t)}{\cos ^m(\chi t)}\right)}
    \qquad , \qquad
    C_{24}=-\frac{1}{4}F(\chi t)\cos^{N-2}{(\chi t)} e^{-\lambda  t \left(1-\frac{\text{sinc}(m \chi t)}{\cos ^m(\chi t)}\right)}\:,
\end{align*}
\begin{align*}
    \Gamma_{11}&=\frac{N-m\lambda t}{4}+\frac{1}{8}F(0)-\frac{1}{8}F(2\chi t)\cos^{N-2}{(2\chi t)} e^{-\lambda  t \left(1-\frac{\text{sinc}(2m \chi t)}{\cos ^m(2\chi t)}\right)},\\
    			  \Gamma_{12}&= \frac{1}{4}F(\chi t) \sin{(\chi t)}\cos^{N-2}{(\chi t)} e^{-\lambda  t \left(1-\frac{\text{sinc}(m \chi t)}{\cos ^m(\chi t)}\right)},\\
    			  \Gamma_{13}&=\frac{1}{16}G(2\chi t)\sin{(2\chi t)}\cos^{N-3}{(2\chi t)} e^{-\lambda  t \left(1-\frac{\text{sinc}(2m \chi t)}{\cos ^m(2\chi t)}\right)},\\
    			  \Gamma_{14}&=\frac{1}{8}F(\chi t)\cos^{N-1}{(\chi t)} e^{-\lambda  t \left(1-\frac{\text{sinc}(m \chi t)}{\cos ^m(\chi t)}\right)}+\frac{1}{32}G(\chi t)\cos^{N-3}{(\chi t)} e^{-\lambda  t \left(1-\frac{\text{sinc}(m \chi t)}{\cos ^m(\chi t)}\right)}-\frac{1}{32}G(3\chi t)\cos^{N-3}{(3\chi t)} e^{-\lambda  t \left(1-\frac{\text{sinc}(3m \chi t)}{\cos ^m(3\chi t)}\right)},\\
     \Gamma_{22}&=\frac{N-m\lambda t}{4},\\
     \Gamma_{23}&=\frac{1}{8}F(\chi t)\cos^{N-1}{(\chi t)} e^{-\lambda  t \left(1-\frac{\text{sinc}(m \chi t)}{\cos ^m(\chi t)}\right)}-\frac{1}{8}G(\chi t)\sin^2{(2\chi t)}\cos^{N-3}{(\chi t)} e^{-\lambda  t \left(1-\frac{\text{sinc}(m \chi t)}{\cos ^m(\chi t)}\right)},\\
    \Gamma_{24}&=\frac{1}{16}G(2\chi t)\sin{(2\chi t)}\cos^{N-3}{(2\chi t)} e^{-\lambda  t \left(1-\frac{\text{sinc}(2m \chi t)}{\cos ^m(2\chi t)}\right)},\\
    \Gamma_{33}&=\frac{3}{32}F(0)-\frac{1}{8}F(2\chi t) \cos^{N-2}{(2\chi t)} e^{-\lambda  t \left(1-\frac{\text{sinc}(2m \chi t)}{\cos ^m(2\chi t)}\right)}+\frac{1}{32}G(0)+\frac{1}{32}G(2\chi t)\cos^{N-2}{(2\chi t)} e^{-\lambda  t \left(1-\frac{\text{sinc}(2m \chi t)}{\cos ^m(2\chi t)}\right)}\\&\quad-\frac{1}{32}I(2\chi t)\sin^2{(2\chi t)}\cos^{N-4}{(2\chi t)} e^{-\lambda  t \left(1-\frac{\text{sinc}(2m \chi t)}{\cos ^m(2\chi t)}\right)},\\
    \Gamma_{34}&=-\frac{1}{16}(F(\chi t)\sin{(\chi t)} \cos^{N-2}{(\chi t)}e^{-\lambda  t \left(1-\frac{\text{sinc}(2m \chi t)}{\cos ^m(2\chi t)}\right)}+F(0))+\frac{1}{16}G(\chi t)\sin{(\chi t)}\cos^{N-2}{(\chi t)}e^{-\lambda  t \left(1-\frac{\text{sinc}(m \chi t)}{\cos ^m(\chi t)}\right)}\\&\quad+\frac{1}{64}I(\chi t)\sin{(\chi t)}\cos^{N-4}{(\chi t)}e^{-\lambda  t \left(1-\frac{\text{sinc}(m \chi t)}{\cos ^m(\chi t)}\right)}+\frac{1}{64}I(3\chi t)\sin{(3\chi t)}\cos^{N-4}{(3\chi t)}e^{-\lambda  t \left(1-\frac{\text{sinc}(3m \chi t)}{\cos ^m(3\chi t)}\right)},\\
\Gamma_{44}&=-\frac{1}{64}F(0)-\frac{1}{128}I(4\chi t)\cos^{N-4}{(4\chi t)}e^{-\lambda  t \left(1-\frac{\text{sinc}(4m \chi t)}{\cos ^m(4\chi t)}\right)}+\frac{1}{128}J(0).
\end{align*}
The functions $F$, $G$, $I$ and $J$ are given by
\begin{align*}
F(\chi t)&=N(N-1)-m(2N-1)\lambda t\frac{\text{sinc}(m \chi t)}{\cos ^m(\chi t)}+m^2\lambda t\frac{\text{sinc}(m \chi t)}{\cos ^m(\chi t)} \left(\lambda  t\frac{\text{sinc}(m \chi t)}{\cos ^m(\chi t)}+1\right),\\
G(\chi t)&=N(N-1)(N-2)-m\left(N(N-1)+(2N-1)(N-2)\right)\lambda t \frac{\text{sinc}(m \chi t)}{\cos ^m(\chi t)}+3m^2(N-1)\lambda t\frac{\text{sinc}(m \chi t)}{\cos ^m(\chi t)}\left(1+\lambda t\frac{\text{sinc}(m \chi t)}{\cos ^m(\chi t)}\right) \\
    					    &\quad-m^3\lambda t\frac{\text{sinc}(m \chi t)}{\cos ^m(\chi t)}\left(1+3\lambda t\frac{\text{sinc}(m \chi t)}{\cos ^m(\chi t)}+(\lambda t\frac{\text{sinc}(m \chi t)}{\cos ^m(\chi t)})^2\right),\\
I(\chi t)&=N(N-1)(N-2)(N-3)-m\left(((N-2) (2N-1)+N(N-1))(N-3)+N(N-1)(N-2)\right)\lambda  t \frac{\text{sinc}(m\chi t)}{\cos ^m(\chi t)}\notag\\&\quad+m^2 (N(N-1)+(2N-1)(N-2)-3(N-1)(N-3))\lambda  t \frac{\text{sinc}(m\chi t)}{\cos ^m(\chi t)}\left(1+\lambda t\frac{\text{sinc}(m\chi t)}{\cos ^m(\chi t)}\right)\\
    & \quad-2m^3(2N-3)\lambda t\frac{\text{sinc}(m\chi t)}{\cos ^m(\chi t)}\left(1+3\lambda t\frac{\text{sinc}(m\chi t)}{\cos ^m(\chi t)}+\left(\lambda t\frac{\text{sinc}(m\chi t)}{\cos ^m(\chi t)}\right)^2\right)\\
    &\quad+m^4\lambda t\frac{\text{sinc}(m\chi t)}{\cos ^m(\chi t)}\left(1+7\lambda t\frac{\text{sinc}(m\chi t)}{\cos ^m(\chi t)}+6\left(\lambda t\frac{\text{sinc}(m\chi t)}{\cos ^m(\chi t)}\right)^2+\left(\lambda t\frac{\text{sinc}(m\chi t)}{\cos ^m(\chi t)}\right)^3\right),\\
    J(\chi t)&=N^2(N-1)(N+3)-2m(N^2(N+1)+N(N-1)(N+3)) \lambda t\frac{\text{sinc}(m\chi t)}{\cos ^m(\chi t)} \notag\\&\quad+m^2(6N^2+2N+1) \lambda t\frac{\text{sinc}(m\chi t)}{\cos ^m(\chi t)}\left(\lambda t\frac{\text{sinc}(m\chi t)}{\cos ^m(\chi t)}+1\right)-2m^3(2N+1)\lambda t\frac{\text{sinc}(m\chi t)}{\cos ^m(\chi t)}\left(1+3\lambda t\frac{\text{sinc}(m\chi t)}{\cos ^m(\chi t)}+\left(\lambda t\frac{\text{sinc}(m\chi t)}{\cos ^m(\chi t)}\right)^2\right)\\
&\quad+m^4\lambda t\frac{\text{sinc}(m\chi t)}{\cos ^m(\chi t)}\left(1+6\lambda t\frac{\text{sinc}(m\chi t)}{\cos ^m(\chi t)}+6\left(\lambda t\frac{\text{sinc}(m\chi t)}{\cos ^m(\chi t)}\right)^2+9\left(\lambda t\frac{\text{sinc}(m\chi t)}{\cos ^m(\chi t)}\right)^3 +3\left(\lambda t\frac{\text{sinc}(m\chi t)}{\cos ^m(\chi t)}\right)^4\right).
\end{align*}



\begin{thebibliography}{}

\bibitem{CavesPRD1981} C. M. Caves, Quantum-mechanical noise in an interferometer, \href{https://doi.org/10.1103/PhysRevD.23.1693}{Phys. Rev. D \textbf{23}, 1693 (1981)}.

\bibitem{WinelandPRA1992} D. J. Wineland, J. J. Bollinger, W. M. Itano, F. L. Moore, and
D. J. Heinzen, Spin squeezing and reduced quantum noise in spectroscopy, \href{https://doi.org/10.1103/PhysRevA.46.R6797}{Phys. Rev. A \textbf{46}, R6797 (1992)}.

\bibitem{BollingerPRA1996}
J. J. Bollinger, W. M. Itano, D. J. Wineland, and D. J. Heinzen, 
Optimal frequency measurements with maximally correlated states,
\href{https://doi.org/10.1103/PhysRevA.54.R4649}{Phys. Rev. A {\bf 54}, R4649 (1996)}.

\bibitem{GiovannettiNATPHOT2011}
V. Giovannetti, S. Lloyd and L. Maccone, 
Advances in quantum metrology, 
\href{https://doi.org/10.1038/nphoton.2011.35}{Nat. Phot. {\bf 5}, 222 (2011)}.

\bibitem{PezzeRMP2018} L. Pezz\`{e}, A. Smerzi, M. K. Oberthaler, R. Schmied, and P. Treutlein, 
Quantum metrology with nonclassical states of atomic ensembles, \href{https://doi.org/10.1103/RevModPhys.90.035005}{Rev. Mod. Phys. \textbf{90}, 035005 (2018)}.

\bibitem{KitagawaPRA1993} M. Kitagawa and M. Ueda, Squeezed spin states, \href{https://doi.org/10.1103/PhysRevA.47.5138}{Phys. Rev. A. \textbf{47}, 5138 (1993)}.

\bibitem{SorensenNATURE2001} A. S\o{}rensen, L. M. Duan, J. I. Cirac, and P. Zoller, Many-particle entanglement with Bose-Einstein condensates, \href{http://dx.doi.org/10.1038/35051038}{Nature \textbf{409}, 63 (2001)}.

\bibitem{EPJD} Y. Li, P. Treutlein, J. Reichel and A. Sinatra, Spin squeezing in a bimodal condensate: spatial dynamics and particle losses, \href{https://doi.org/10.1140/epjb/e2008-00472-6}{Eur. Phys. J. B {\bf 68}, 365 (2009)}.

\bibitem{GrossNATURE2010} C. Gross, T. Zibold, E. Nicklas, J. Est\`{e}ve, and M. K. Oberthaler, Nonlinear atom interferometer surpasses classical precision limit, \href{https://doi.org/10.1038/nature08919}{Nature \textbf{464}, 1165 (2010)}.

\bibitem{Treutlein2010} M. F. Reidel, P. B\"{o}hi, Y. Li, T. W. H\"{a}nsch, A. Sinatra, P. Treutlein, Atom chip based generation of entanglement for quantum metrology, \href{https://doi.org/10.1038/nature08988}{Nature {\bf 464}, 1170 (2010)}.

\bibitem{LerouxPRL2010}
I. D. Leroux, M. H. Schleier-Smith, and V. Vuleti\`c, 
Implementation of cavity squeezing of a collective atomic spin, 
\href{https://doi.org/10.1103/PhysRevLett.104.073602}{Phys. Rev. Lett. {\bf 104}, 073602 (2010)}.

\bibitem{MitchellPRL2012} R. J. Sewell, M. Koschorreck, M. Napolitano, B. Dubost, N. Behbood, and M. W. Mitchell, Magnetic Sensitivity Beyond the Projection Noise Limit by Spin Squeezing, \href{https://doi.org/10.1103/PhysRevLett.109.253605}{Phys. Rev. Lett. \textbf{109}, 253605 (2012)}.

\bibitem{HostenNATURE2016} O. Hosten, N. J. Engelsen, R. Krishnakumar and M. A. Kasevich, Measurement noise 100 times lower than the quantum-projection limit using entangled atoms, \href{https://doi.org/10.1038/nature16176}{Nature \textbf{529}, 505 (2016)};

\bibitem{CoxPRL2016} 
K. C. Cox, G. P. Greve, J. M. Weiner, and J. K. Thompson, 
Deterministic squeezed states with collective measurements and feedback, \href{https://doi.org/10.1103/PhysRevLett.116.093602}{Phys. Rev. Lett. {\bf 116}, 093602 (2016)}.

\bibitem{ChalopinNCom2018}
T. Chalopin, C. Bouazza, A. Evrard, V. Makhalov, D. Dreon, J. Dalibard, L. A. Sidorenkov, S. Nascimbene,
Quantum-enhanced sensing using non-classical spin states of a highly magnetic atom, \href{https://www.nature.com/articles/s41467-018-07433-1}{Nat. Commun. {\bf 9}, 4955 (2018).}

\bibitem{VuleticArxiv2021} S. Colombo, E. Pedrozo-Pe\~nafiel, A. F. Adiyatullin, Z. Li, E. Mendez, C. Shu and V. Vuleti\`c, Time-Reversal-Based Quantum Metrology with Many-Body Entangled States, \href{https://arxiv.org/abs/2106.03754}{arXiv:2106.03754}.

%
%

%

\bibitem{YurkePRA1986} B. Yurke, S. L. McCall, and J. R. Klauder, SU(2) and SU(1,1) interferometers, \href{https://doi.org/10.1103/PhysRevA.33.4033}{Phys. Rev. A \textbf{33}, 4033 (1986)}.

\bibitem{SchleierSmithPRL2016} E. Davis, G. Bentsen and M. Schleier-Smith, Approaching the Heisenberg Limit without Single-Particle Detection, \href{https://doi.org/10.1103/PhysRevLett.116.053601}{Phys. Rev. Lett. \textbf{116}, 053601 (2016)}.

\bibitem{FrowisPRL2016} F. Fr\"{o}wis, P. Sekatski, and Wolfgang D\"{u}r, Detecting Large Quantum Fisher Information with Finite Measurement Precision, \href{https://doi.org/10.1103/PhysRevLett.116.090801}{Phys. Rev. Lett. \textbf{116}, 090801 (2016)}.

\bibitem{MacriPRA2016} T. Macr\`{i}, A. Smerzi, and L. Pezz\`{e}, Loschmidt echo for quantum metrology, \href{https://doi.org/10.1103/PhysRevA.94.010102}{Phys. Rev. A \textbf{94}, 010102(R) (2016)}.

\bibitem{HostenSCIENCE2016} O. Hosten, R. Krishnakumar, N. J. Engelsen, and M. A. Kasevich, Quantum phase magnification, \href{https://doi.org/10.1126/science.aaf3397}{Science \textbf{352}, 1552 (2016)}.

\bibitem{NolanPRL2017} S. P. Nolan, S. S. Szigeti, and S. A. Haine, Optimal and Robust Quantum Metrology Using Interaction-Based Readouts, \href{https://doi.org/10.1103/PhysRevLett.119.193601}{Phys. Rev. Lett. \textbf{119}, 193601 (2017)}.

\bibitem{HainePRA2018}
S. A. Haine, Using interaction-based readouts to approach the ultimate limit of detection-noise robustness for quantum-enhanced metrology in collective spin systems, \href{https://doi.org/10.1103/PhysRevA.98.030303}{Phys. Rev. A \textbf{98}, 030303(R) (2018)}.

\bibitem{HammererQuantum2020} M. Schulte, V. J. Martínez-Lahuerta, M. S. Scharnagl and K. Hammerer, Ramsey interferometry with generalized one-axis twisting echoes, \href{https://doi.org/10.22331/q-2020-05-15-268}{Quantum. \textbf{4}, 268 (2020)}.

\bibitem{HuelgaPRL1997} S. F. Huelga, C. Macchiavello, T. Pellizzari, A. K. Ekert, M. B. Plenio, and J. I. Cirac, Improvement of Frequency Standards with Quantum Entanglement, \href{https://doi.org/10.1103/PhysRevLett.79.3865}{Phys. Rev. Lett. \textbf{79}, 3865 (1997)}.

\bibitem{MonzPRL2011}
T. Monz, P. Schindler, J. T. Barreiro, M. Chwalla, D. Nigg, W. A. Coish, M. Harlander, W. H\"ansel, M. Hennrich, and R. Blatt, 
14-qubit entanglement: Creation and coherence, 
\href{https://doi.org/10.1103/PhysRevLett.106.130506}{Phys. Rev. Lett. {\bf 106}, 130506 (2011)}.

\bibitem{DemkowiczNATCOMMUN2012} R. Demkowicz-Dobrza\'{n}ski, J. Ko\l{}ody\'{n}ski, and M. Gu\c{t}\v{a}, The elusive Heisenberg limit in quantum-enhanced metrology, \href{https://doi.org/10.1038/ncomms2067}{Nat. Commun. \textbf{3}, 1063 (2012)}.

\bibitem{Kittens} K. Pawlowski, M. Fadel, P. Treutlein, Y. Castin, and A. Sinatra, Mesoscopic quantum superpositions in bimodal Bose-Einstein condensates: Decoherence and strategies to counteract it, \href{https://doi.org/10.1103/PhysRevA.95.063609}{Phys. Rev. A {\bf 95}, 063609 (2017)}.


\bibitem{StrobelSCIENCE2014}
H. Strobel, W. Muessel, D. Linnemann, T. Zibold, D. B. Hume, L. Pezz\`e, A. Smerzi, and M. K. Oberthaler, 
Fisher information and entanglement of non-Gaussian spin states, 
\href{https://doi.org/10.1126/science.1250147 }{Science {\bf 345}, 424 (2014)}.

\bibitem{BohnetSCIENCE2016} J. G. Bohnet, B. C. Sawyer, J. W. Britton, M. L. Wall, A. M. Rey, M. Foss-Feig, J. J. Bollinger, Quantum spin dynamics and entanglement generation with hundreds of trapped ions, \href{https://doi.org/10.1126/science.aad9958 }{Science \textbf{352}, 1297 (2016)}.

\bibitem{EvrardPRL2019} A. Evrard, V. Makhalov, T. Chalopin, L. A. Sidorenkov, J. Dalibard, R. Lopes, and S. Nascimbene, Enhanced Magnetic Sensitivity with Non-Gaussian Quantum Fluctuations, \href{https://doi.org/10.1103/PhysRevLett.122.173601}{Phys. Rev. Lett. \textbf{122}, 173601 (2019)}.

\bibitem{XuArXiv2021} K. Xu, Y.-R. Zhang, Z.-H. Sun, H. Li, P. Song, Z. Xiang, K. Huang, H. Li, Y.-H. Shi, C.-T. Chen, X. Song, D. Zheng, F. Nori, H. Wang, H. Fan, Metrological characterisation of non-Gaussian entangled states of superconducting qubits, \href{https://arxiv.org/abs/2103.11434}{arXiv:2103.11434}.

\bibitem{GessnerPRL2019} M. Gessner, A. Smerzi and L. Pezz\`{e}, Metrological Nonlinear Squeezing Parameter, \href{https://doi.org/10.1103/PhysRevLett.122.090503}{Phys. Rev. Lett. \textbf{122}, 090503 (2019)}.

\bibitem{BaamaraPRL2021} Y. Baamara, A. Sinatra, M. Gessner, Scaling laws for the sensitivity enhancement of non-Gaussian spin states, \href{https://doi.org/10.1103/PhysRevLett.127.160501}{Phys. Rev. Lett. \textbf{127}, 160501 (2021)}.

\bibitem{SinatraFro2012} A. Sinatra, J.-C. Dornstetter and Y. Castin, Spin squeezing in Bose-Einstein condensates: Limits imposed by decoherence and non-zero temperature, \href{https://doi.org/10.1007/s11467-011-0219-7}{Front. Phys. \textbf{7}, 86 (2012)}.

\bibitem{LuckeSCIENCE2011}
B. L\"{u}cke, M. Scherer, J. Kruse, L. Pezz\`{e}, F. Deuretzbacher, P. Hyllus, O. Topic, J. Peise, W. Ertmer, J. Arlt, L. Santos, A. Smerzi, and C. Klempt, Twin Matter Waves for Interferometry Beyond the Classical Limit, \href{http://dx.doi.org/10.1126/science.1208798}{Science \textbf{334}, 773 (2011)}.

\bibitem{BraunsteinPRL1994} S. L. Braunstein and C. M. Caves, Statistical distance and the geometry of quantum states, \href{https://doi.org/10.1103/PhysRevLett.72.3439}{Phys. Rev. Lett. \textbf{72}, 3439 (1994)}.

\bibitem{LiYunPRL2008} Y. Li, Y. Castin, and A. Sinatra, Optimum Spin Squeezing in Bose-Einstein Condensates with Particle Losses, \href{https://doi.org/10.1103/PhysRevLett.100.210401}{Phys. Rev. Lett. \textbf{100}, 210401 (2008)}.

\bibitem{SinatraPRL2011} A. Sinatra, E. Witkowska, J.-C. Dornstetter, Y. Li, and Y. Castin, Limit of Spin Squeezing in Finite-Temperature Bose-Einstein Condensates, \href{https://doi.org/10.1103/PhysRevLett.107.060404}{Phys. Rev. Lett. \textbf{107}, 060404 (2011)}.

\bibitem{MolmerPRL1999} K. M\o{}lmer and A. S\o{}rensen, Multiparticle Entanglement of Hot Trapped Ions, \href{https://doi.org/10.1103/PhysRevLett.82.1835}{Phys. Rev. Lett. \textbf{82}, 1835 (1999)}.

\bibitem{LanyonPRL2013} B. P. Lanyon, P. Jurcevic, C. Hempel, M. Gessner, V. Vedral, R. Blatt, and C. F. Roos, Experimental Generation of Quantum Discord via Noisy Processes, \href{https://doi.org/10.1103/PhysRevLett.111.100504}{Phys. Rev. Lett. \textbf{111}, 100504 (2013)}.

\bibitem{CarnioPRL2015} E. G. Carnio, A. Buchleitner, and M. Gessner, Robust Asymptotic Entanglement under Multipartite Collective Dephasing, \href{https://doi.org/10.1103/PhysRevLett.115.010404}{Phys. Rev. Lett. \textbf{115}, 010404 (2015)}.

\bibitem{LerouxPRA2012} I. D. Leroux, M. H. Schleier-Smith, H. Zhang, and V. Vuleti\'c,
Unitary cavity spin squeezing by quantum erasure, 
\href{https://doi.org/10.1103/PhysRevA.85.013803}{Phys. Rev. A \textbf{85}, 013803 (2012)}.

\bibitem{PawlowskiEPL2016} K. Pawłowski, J. Estève, J. Reichel and A. Sinatra, Limits of atomic entanglement by cavity feedback: From weak to strong coupling, \href{https://doi.org/10.1209/0295-5075/113/34005}{EPL \textbf{113}, 34005 (2016)}.

\bibitem{Tannoudji} Claude Cohen-Tannoudji, Jacques Dupont-Roc, Gilbert Grynberg, \textit{Atom—Photon Interactions: Basic Process and Appilcations}, (WILEY‐VCH Verlag GmbH $\&$ Co. KGaA, 2004).

\bibitem{Breuer} H.-P. Breuer and F. Petruccione, \textit{The Theory of Open Quantum Systems}, (Oxford University Press, Oxford, U.K., 2007).

\bibitem{MonikaPRA2010} M. H. Schleier-Smith, I. D. Leroux, and V. Vuleti\'{c}. Squeezing the collective spin of a dilute atomic ensemble by cavity feedback, \href{https://doi.org/110.1103/PhysRevA.83.039907}{Phys. Rev. A {\bf 81}, 021804, (2010)}.


\bibitem{SinatraEuro1998} A. Sinatra, and Y. Castin. Phase dynamics of Bose-Einstein condensates: Losses versus revivals, \href{https://doi.org/10.1007/s100530050206}{Eur. Phys. J. D 4, 247–260 (1998)}.

\bibitem{MolmerOpt1993} K. Mølmer, Y. Castin, and J. Dalibard. A Monte-Carlo wave function method in quantum optics, \href{https://doi.org/10.1364/JOSAB.10.000524}{J. Opt. Soc. Am. B, 10:524, 1993}.

%

%








\end{thebibliography}
\end{document}